\documentclass[prd,showpacs,%twocolumn,
superscriptaddress,preprintnumbers,nofootinbib,floatfix]{revtex4}

\usepackage{amsmath,graphicx,amsfonts,array,amsthm,bm}
\begin{document}

\preprint{KU-TP 037}

\title{Global solutions for higher-dimensional stretched
small black holes}

\author{Chiang-Mei Chen} \email{cmchen@phy.ncu.edu.tw}
\affiliation{Department of Physics and Center for Mathematics and Theoretical
Physics, National Central University, Chungli 320, Taiwan}

\author{Dmitri V. Gal'tsov} \email{galtsov@phys.msu.ru}
\affiliation{Department of Theoretical Physics, Moscow State University,
119899, Moscow, Russia}

\author{Nobuyoshi Ohta} \email{ohtan@phys.kindai.ac.jp}
\affiliation{Department of Physics, Kinki University, Higashi-Osaka, Osaka 577-8502, Japan}

\author{Dmitry G. Orlov} \email{orlov_d@mail.ru}
\affiliation{Department of Physics, National Central University,Chungli 320, Taiwan}

\date{\today}

\begin{abstract}
Small black holes in heterotic string theory have vanishing horizon
area at the supergravity level, but the horizon is stretched to the
finite radius $AdS_2 \times S^{D-2}$ geometry once higher curvature
corrections are turned on. This has been demonstrated to give good
agreement with microscopic entropy counting. Previous
considerations, however, were based on the classical local solutions
valid only in the vicinity of the event horizon. Here we address the
question of global existence of extremal black holes in the
$D$-dimensional Einstein-Maxwell-Dilaton theory with the
Gauss-Bonnet term introducing a variable dilaton coupling $a$ as a
parameter. We show that asymptotically flat black holes exist only
in a bounded region of the dilaton couplings $0 < a < a_{\rm cr}$ where
$a_{\rm cr}$ depends on $D$. For $D \geq 5$ (but not for $D = 4$) the
allowed range of $a$ includes the heterotic string values. For
$a > a_{\rm cr}$ numerical solutions meet weak naked singularities at
finite radii $r = r_{\rm cusp}$ (spherical cusps), where the scalar
curvature diverges as $|r - r_{\rm cusp}|^{-1/2}$. For $D \geq 7$ cusps
are met in pairs, so that solutions can be formally extended to
asymptotically flat infinity choosing a suitable integration
variable. We show, however, that radial geodesics cannot be
continued through the cusp singularities, so such a continuation is
unphysical.
\end{abstract}

%% REVTEX4

\maketitle

%%%%%%%%%%%%%%%%%%%%%%%%%%%%%%%%%%%%%%%%%%%%%%%%%%%%%%%%%%%%%%%%%%%%%%
%%%%%%%%%%%%%%%%%%%%%%%%%%%%%%%%%%%%%%%%%%%%%%%%%%%%%%%%%%%%%%%%%%%%%%
\section{Introduction}
\label{intro}
%%%%%%%%%%%%%%%%%%%%%%%%%%%%%%%%%%%%%%%%%%%%%%%%%%%%%%%%%%%%%%%%%%%%%%
%%%%%%%%%%%%%%%%%%%%%%%%%%%%%%%%%%%%%%%%%%%%%%%%%%%%%%%%%%%%%%%%%%%%%%
During recent years important progress has been achieved in
understanding the entropy of the so-called small black holes (for
reviews see~\cite{deWit:2005ya, Mohaupt:2005jd, Sen:2007qy}) which
have vanishing horizon area at the supergravity level~\cite{Gibbons:1982ih,
Gibbons:1985ac, Gibbons:1987ps, Garfinkle:1990qj}.
The discrepancy with the microscopic counting which gives the finite
entropy was resolved by the discovery that the area of the horizon
is stretched to finite radius once curvature corrections are
included. Indeed such corrections have long been known to exist in
the low-energy effective theories of
superstrings~\cite{Zwiebach:1985uq, Callan:1986jb, Gross:1986iv,
Metsaev:1987zx, Gross:1986mw}. The classically computed entropy then differs
from the Bekenstein-Hawking value~\cite{Wald:1993nt,
Jacobson:1993vj, Iyer:1994ys, Jacobson:1994qe}, but agrees (at least
up to a coefficient) with microscopic counting~\cite{Behrndt:1998eq,
LopesCardoso:1998wt, LopesCardoso:1999cv, LopesCardoso:1999ur,
LopesCardoso:1999xn, Mohaupt:2000mj, LopesCardoso:2000qm,
LopesCardoso:2000fp, Dabholkar:2004yr, Dabholkar:2004dq, Sen:2004dp,
Hubeny:2004ji, Bak:2005mt} including non-BPS
cases~\cite{Goldstein:2005hq, Kallosh:2005ax, Tripathy:2005qp,
Giryavets:2005nf, Goldstein:2005rr, Kallosh:2006bt, Kallosh:2006bx,
Prester:2005qs,Cvitan:2007pk,Cvitan:2007en,Prester:2008iu,
Alishahiha:2006ke, Sinha:2006yy, Chandrasekhar:2006kx,
Parvizi:2006uz, Sahoo:2006rp, Astefanesei:2006sy}. Within the models
in which the supersymmetric versions of the curvature square terms
are available, the correspondence was checked using the exact
classical solutions~\cite{Dabholkar:2004yr, Dabholkar:2004dq,
Bak:2005mt}. It was also observed that good agreement is achieved if
the curvature corrections are taken in the form of the Gauss-Bonnet
(GB) term both in 4D and higher dimensions~\cite{Prester:2005qs,
Cvitan:2007pk, Cvitan:2007en, Prester:2008iu}.

To compute the entropy of extremal black holes with the horizon
$AdS_2 \times S^{D-2}$ from the classical side it is enough to
construct local solutions in the vicinity of the horizon which is
easily done analytically~\cite{Sen:2005wa, Sen:2005iz, Cai:2007cz}.
But this does not guarantee the existence of global asymptotically
flat solutions. Construction of solutions with curvature
corrections, apart from purely perturbative probes~\cite{Callan:1988hs,
Mignemi:1992nt, Mignemi:1993ce}, requires numerical integration of
the field equations. For non-extremal black holes this was done
in~\cite{Kanti:1995vq,Torii:1996yi,Alexeev:1997ua, Melis:2005xt,
Melis:2005ji, Alexeev:1996vs, Guo:2008hf, Guo:2008eq, Ohta:2009tb,
Ohta:2009pe, Maeda:2009uy}.
The global existence of extremal black holes in the 4D model with
the GB terms endowed with an arbitrary dilaton coupling $a$ was
proven in~\cite{Chen:2006ge,Chen:2008px}. It turned out that global
asymptotically flat black holes with the horizon $AdS_2 \times S^2$
existed for the dilaton coupling below the critical value of the
order $a_{\rm cr} \sim 1/2$ and less than $\frac12$. This range does
not include the heterotic string value $a = 1$ nor $\frac12$.
This result is modified in the presence of the magnetic
charge~\cite{Chen:2008hk}, which extends
the region of the allowed couplings and serve as the order parameter
ensuring continuous transition to the theory without curvature
corrections. It is worth noting that our model has neither
continuous nor discrete S-duality, so properties of the purely
electric solution essentially differs from that of dyons.

The purpose of the present paper is to investigate existence of
global solutions for small stretched purely electric black holes in
higher dimensional Einstein-Maxwell-Dilaton theory with the
Gauss-Bonnet term (EMDGB). We construct the local solutions in terms
of series expansion around the degenerate event horizon for an
arbitrary space-time dimension $D$ and calculate the discrete
sequence of black hole entropies using Sen's entropy function
approach. The entropy is found to be monotonically increasing with
$D$. Then we continue numerically these local solutions and show
that in dimensions higher than four the heterotic string value of
the dilaton coupling lies inside the range of the existence of global
asymptotically flat static black holes.

We also investigate physical significance of the so-called turning
points which were encountered in numerical solutions within the
four-dimensional EMDGB theory~\cite{Kanti:1995vq, Torii:1996yi,
Alexeev:1996vs, Alexeev:1997ua, Chen:2006ge, Chen:2008hk}.
They correspond to mild singularities at finite radii outside the
horizon where metric and its first derivatives are finite, but the
second derivatives diverge. Numerical solutions can be extended
through these singularities, which we call `cusps' in this paper, by
suitable redefinition of the integration
variable~\cite{Alexeev:1997ua, Pomazanov:2000}. In four dimensions the
solution extended this way then meets a stronger singularity at
finite distance, so actually the cusp is just a precursor of the
strong singularity. In higher dimensions ($D \ge 7$) we encounter an
interesting new feature: the cusps come out in pairs of right and
left turning points, so the extended solution finally may be even
asymptotically flat. This could correspond to a novel type of black
hole coated by cusp pairs. But somewhat disappointingly, our
analysis shows that continuation of geodesics through the cusp
singularities in the extended manifolds cannot be performed in a
smooth way. Thus we are inclined to reject such extended manifolds
as physical black hole solutions. Instead, we interpret the
occurrence of cusp singularity as failure to produce asymptotically
flat black holes. This gives an upper bound on the dilaton coupling.
We find numerically the sequence of critical dilaton couplings for
$4 \leq D \leq 10$ which turns out to be increasing with $D$.

Another novel feature of EMDGB black holes with a degenerate horizon
in higher dimensions is that the role of the GB term in the near
critical solutions may still be significant. In four dimensions, as
was shown in~\cite{Chen:2006ge}, the near-critical solutions
saturate the BPS bounds of the corresponding theory without
curvature corrections. This means that relative contribution of
the GB term becomes negligible when the dilaton coupling approaches
its upper boundary. We find that for $D \ge 7$ this is not so, and
the BPS conditions are not satisfied in this limit.

This paper is organized as follows. In Sec.~II, we define the
action, present the field equations in various forms and discuss
symmetries of the system. In Sec.~III, we review solutions for small
black holes without GB corrections as well and the solutions with
the GB term but without dilaton. Then we construct the local series
solutions near the horizon and calculate the entropy of stretched
black holes using Sen's entropy function. We obtain the discrete
sequence of the entropies of curvature corrected black holes in
various dimensions interpolating starting with twice the
Hawking-Bekenstein value $A/2$ for $D =4$ up to $41A/52$ for $D = 10$.
In Sec.IV, we present asymptotic expansions of the desired
solutions, introduce global charges  and discuss the BPS conditions.
The next Sec.~V is devoted to the cusp problem. We explain why
extension of solutions through the cusp singularity is physically
unacceptable. Finally in Sec.~VI we present numerical results for
various dimensions and explore the fulfillment of the BPS conditions
on the boundary of the allowed dilaton couplings.

%%%%%%%%%%%%%%%%%%%%%%%%%%%%%%%%%%%%%%%%%%%%%%%%%%%%%%%%%%%%%%%%%%%%%%
%%%%%%%%%%%%%%%%%%%%%%%%%%%%%%%%%%%%%%%%%%%%%%%%%%%%%%%%%%%%%%%%%%%%%%
\section{Setup}
\label{setup}
%%%%%%%%%%%%%%%%%%%%%%%%%%%%%%%%%%%%%%%%%%%%%%%%%%%%%%%%%%%%%%%%%%%%%%
%%%%%%%%%%%%%%%%%%%%%%%%%%%%%%%%%%%%%%%%%%%%%%%%%%%%%%%%%%%%%%%%%%%%%%
A low-energy bosonic effective action for the heterotic string theory
with the curvature corrections is given by~\cite{Metsaev:1987zx,
Gross:1986mw}
\begin{equation}
I = \frac1{16 \pi G} \int d^Dx \sqrt{-\tilde g} \Phi \left( \tilde R
+ \Phi^{-2} \tilde \partial_\mu \Phi \tilde \partial^\mu \Phi
- \tilde F_{\mu\nu} \tilde F^{\mu\nu} + \frac{\alpha'}8 \tilde{\cal L}_{\rm GB} \right),
\end{equation}
where $\tilde F^{\mu\nu}$ is the Maxwell field (we use a truncation
involving only one $U(1)$ field), $\tilde{{\cal L}}_{\rm GB}$ is
the Euler density
\begin{equation}
\tilde{{\cal L}}_{\rm GB} = \tilde{R}^2 - 4 \tilde{R}_{\mu\nu} \tilde{R}^{\mu\nu} +
\tilde{R}_{\alpha\beta\mu\nu} \tilde{R}^{\alpha\beta\mu\nu},
\end{equation}
and $\alpha'$ is the Regge slope parameter.
The tilde denotes the quantities related to the string frame
metric $\tilde g_{\mu\nu}$. The action can be transformed to the
Einstein frame with metric $g_{\mu\nu}$ by the conformal rescaling
\begin{equation}
g_{\mu\nu} = \Phi^{\frac2{D-2}} \, \tilde g_{\mu\nu},
\end{equation}
giving
\begin{equation}
I = \frac1{16 \pi G} \int d^Dx \sqrt{- g} \left( R - \frac{\Phi^{-2}}{D-2}
\partial_\mu \Phi \partial^\mu \Phi - \Phi^{\frac2{D-2}} F_{\mu\nu} F^{\mu\nu}
+ \frac{\alpha'}8 \Phi^{\frac2{D-2}} {\cal L}_{\rm GB}
+ \mathcal{F}(\partial \Phi, R) \right),
\end{equation}
where $\mathcal{F}(\partial \Phi, R)$ denotes the cross terms of
$\partial \Phi$ and curvature coming from the GB term under the
frame transformation. For simplicity, we do not include these terms
in our analysis. We expect that inclusion of these terms might
affect the black hole properties only quantitatively but not qualitatively.
Then redefining the dilaton field as
\begin{equation}
\Phi = \mathrm{e}^{\sqrt{2(D-2)} \; \phi},
\end{equation}
we obtain the action
\begin{equation}
I = \frac1{16 \pi G} \int d^Dx \sqrt{-g} \left( R - 2 \partial_\mu \phi \partial^\mu \phi
- {\rm e}^{2 \sqrt{2/(D-2)} \, \phi} F_{\mu\nu} F^{\mu\nu} +
\frac{\alpha'}8 {\rm e}^{2 \sqrt{2/(D-2)} \, \phi} {\cal L}_{\rm GB} \right).
\end{equation}
In this action we have the sequence of the dilaton couplings
\begin{equation}\label{astring}
a_{\rm str}^2 = \frac2{D-2},
\end{equation}
relevant for the string theory. If we do this in 4 dimensions, we have the
dilaton coupling $a_{\rm str} = 1$, but if we do this in 10 dimensions, we have
$a_{\rm str} = 1/2$. It will be convenient, however, to
consider the above action for two arbitrary dilaton couplings
$a$ and $b$:
\begin{equation} \label{action}
I = \frac1{16 \pi G} \int
d^Dx \sqrt{-g} \left( R - 2 \partial_\mu \phi \partial^\mu \phi -
{\rm e}^{2 a \phi} F_{\mu\nu} F^{\mu\nu} + \alpha {\rm e}^{2 b \phi}
{\cal L}_{\rm GB} \right),
\end{equation}
where we also denoted the GB coupling $\alpha'/8 = \alpha$.

The space-time metric is parametrized by two functions $\omega(r)$ and
$\rho(r)$:
\begin{equation}\label{met}
ds^2 = - \omega(r) dt^2 + \frac{dr^2}{\omega(r)} + \rho^2(r)
d\Omega_{D-2}^2.
\end{equation}
For convenience, we list in Appendix A the relevant geometric
quantities for more general static spherically symmetric metrics.

We will consider only purely electric static spherically symmetric
configurations of the $D$-dimensional Maxwell field
\begin{equation}
A = - f(r) \, dt.
\end{equation}
Then, integrating the Maxwell equations
\begin{equation}
\left( \rho^{D-2} f'
{\rm e}^{2 a \phi} \right)' = 0,
\end{equation}
one obtains
\begin{equation}\label{Solf}
f'(r) = q_e \rho^{2-D} {\rm e}^{- 2 a \phi},
\end{equation}
where $q_e$ is the electric charge, which is considered as a free
parameter (note that the physical electric charge defined
asymptotically differs from this quantity, see Sec.~\ref{global}).

%%%%%%%%%%%%%%%%%%%%%%%%%%%%%%%%%%%%%%%%%%%%%%%%%%%%%%%%%%%%%%%%%%%%%%
\subsection{Field equations}
\label{field}
%%%%%%%%%%%%%%%%%%%%%%%%%%%%%%%%%%%%%%%%%%%%%%%%%%%%%%%%%%%%%%%%%%%%%%
We present the Einstein equations in the form
\begin{equation}
G_{\mu\nu} = 8 \pi G ( T_{\mu\nu}^{\rm mat} + T_{\mu\nu}^{\rm GB} ),
\end{equation}
where
$T_{\mu\nu}^{\rm mat}$ is the matter stress-tensor
\begin{equation}
8 \pi G \, T_{\mu\nu}^{\rm mat} = 2 \left[ \partial_\mu \phi \partial_\nu \phi
- \frac12 \partial_\alpha \phi \partial^\alpha \phi \, g_{\mu\nu}
+ {\rm e}^{2 a \phi} \left( F_{\mu\alpha} F_\nu{}^\alpha
- \frac14 F_{\alpha\beta} F^{\alpha\beta} \, g_{\mu\nu} \right) \right],
\end{equation}
and the $T_{\mu\nu}^{\rm GB}$ is the effective gravitational stresses due
to the GB term
\begin{equation}
8 \pi G \, T_{\mu\nu}^{\rm GB} = - \alpha {\rm e}^{2 b \phi} \left[ H_{\mu\nu}
+ 8 \left( 2 b^2 \nabla^\alpha \phi \nabla^\beta \phi + b \nabla^\alpha
\nabla^\beta \phi \right) P_{\mu\alpha\nu\beta} \right],
\end{equation}
where
\begin{eqnarray}
H_{\mu\nu} &=& 2 ( R R_{\mu\nu} - 2 R_{\mu\alpha} R^\alpha{}_\nu
- 2 R^{\alpha\beta} R_{\mu\alpha\nu\beta} + R_{\mu\alpha\beta\gamma}
R_\nu{}^{\alpha\beta\gamma} ) - \frac12 {\cal L}_{GB} \; g_{\mu\nu},
\\
P_{\mu\alpha\nu\beta} &=& R_{\mu\alpha\nu\beta} + 2 g_{\mu[\beta} R_{\nu]\alpha}
+ 2 g_{\alpha[\nu} R_{\beta]\mu} + R g_{\mu[\nu} g_{\beta]\alpha}.
\end{eqnarray}
For the metric (\ref{met}), the components of $G_{\mu\nu}$ are
\begin{eqnarray}
G_{tt} &=& - \frac{(D-2) \omega}{2 \rho^2} \left[ 2 \omega \rho \rho''
+ \rho \omega' \rho' + (D - 3) (\omega \rho'^2 - 1) \right],
\nonumber\\
G_{rr} &=& \frac{D-2}{2 \omega \rho^2} \left[ \rho \omega' \rho'
+ (D - 3) (\omega \rho'^2 - 1) \right],
\nonumber\\
G_{\theta\theta} &=& \frac12 \rho^2 \omega'' + \frac{D-3}2
\left[ 2 \omega \rho \rho'' + 2 \rho \omega' \rho' + (D - 4) (\omega \rho'^2 - 1) \right],
\end{eqnarray}
while the energy-momentum due to matter fields is given by
\begin{eqnarray}
8 \pi G T_{tt}^{\rm mat} &=& \omega^2 \phi'^2 + {\rm e}^{2 a \phi} \omega f'^2,
\nonumber\\
8 \pi G T_{rr}^{\rm mat} &=& \phi'^2 - {\rm e}^{2 a \phi} \frac{f'^2}{\omega},
\nonumber\\
8 \pi G T_{\theta\theta}^{\rm mat} &=& - \rho^2 \left( \omega
\phi'^2 - {\rm e}^{2 a \phi} f'^2 \right).
\end{eqnarray}
The energy-stress tensor due to the GB term is more complicated
\begin{eqnarray}
\frac{8 \pi G}{\alpha \mathrm{e}^{2 b \phi}} T^{GB}_{tt}
&=& - \frac{D^2_4 \omega}{\rho^3} (2 \omega \rho'' + \omega' \rho')
(\omega \rho'^2 - 1) - \frac{D^2_5 \omega}{2 \rho^4} (\omega \rho'^2 - 1)^2
\nonumber\\
&& - \frac{2 b D^2_3 \omega}{\rho^2} \left[ (2 \omega \phi'' + \omega' \phi')
(\omega \rho'^2 - 1) + 2 \omega \rho' \phi' (2 \omega \rho'' + \omega' \rho') \right]
\nonumber\\
&& - \frac{4 b D^2_4 \omega^2 \rho' \phi'}{\rho^3} (\omega \rho'^2 - 1)
- \frac{8 b^2 D^2_3 \omega^2 \phi'^2}{\rho^2} (\omega \rho'^2 - 1),
\nonumber\\
\frac{8 \pi G}{\alpha \mathrm{e}^{2 b \phi}} T^{GB}_{rr}
&=& \frac{D^2_4 \omega' \rho'}{\omega \rho^3} (\omega \rho'^2 - 1)
+ \frac{D^2_5}{2 \omega \rho^4} (\omega \rho'^2 - 1)^2
\nonumber\\
&& + 2 b \left[ \frac{D^2_3 \omega' \phi'}{\omega \rho^2}
(3 \omega \rho'^2 - 1) + \frac{2 D^2_4 \rho' \phi'}{\rho^3}
(\omega \rho'^2 - 1) \right],
\nonumber\\
\frac{8 \pi G}{\alpha \mathrm{e}^{2 b \phi}} T^{GB}_{\theta\theta}
&=& D^3_4 [ \omega'' (\omega \rho'^2 - 1) + 2 \omega \omega' \rho' \rho''
+ \omega'^2 \rho'^2] + \frac{2 D^3_5}{\rho} (\omega \rho')' (\omega \rho'^2 - 1)
+ \frac{D^3_6}{2 \rho^2} (\omega \rho'^2 - 1)^2
\nonumber\\
&& + 4 b \Biggl[ D^3_3 \rho (\omega \omega' \rho' \phi')' + D^3_4
(\omega \phi')' (\omega \rho'^2 - 1) + 2 D^3_4 \omega \rho' \phi'
(\omega \rho')' \nonumber\\ && + \frac{D^3_5 \omega \rho'
\phi'}{\rho} (\omega \rho'^2 - 1) \Biggr] + 8 b^2 \left[ D^3_3
\omega \omega' \rho \rho' \phi'^2 + D^3_4 \omega \phi'^2 (\omega
\rho'^2 - 1) \right],
\end{eqnarray}
where we have introduced the dimension-dependent coefficients
\begin{equation}
D^m_n = (D - m)_n = (D - m) (D - m - 1) \cdots (D - n), \qquad n \ge m.
\end{equation}

The dilaton equation reads
\begin{eqnarray}
2 (\omega \phi')' + 2 D^2_2 \omega \phi' \frac{\rho'}{\rho}
+ 2 a f'^2 {\rm e}^{2 a \phi}
%\nonumber\\
+ \alpha b D^2_3 {\rm e}^{2 b \phi} \left\{ 2 \frac{[\omega' (\omega
\rho'^2 - 1)]'}{\rho^2} + D^4_5 \frac{(\omega \rho'^2 -
1)^2}{\rho^4} + 4 D^4_4  (\omega \rho')' \frac{\omega \rho'^2 -
1}{\rho^3} \right\} = 0.
\end{eqnarray}

{}From the Einstein equation, one can derive the following two
second order equations for the metric functions $\rho(r)$ and
$\omega(r)$,~\footnote{Namely, the equation for $\rho$ is $-
\frac{\rho^2}{\omega^2} \left[  (\mbox{Einstein equation})_{tt} +
\omega^2 (\mbox{Einstein equation})_{rr} \right]$ and the equation
for $\omega$ is $\frac2{\rho} (\mbox{Einstein
equation})_{\theta\theta}$.} which are more convenient for
numerical integration:
\begin{equation}
D^2_2 \rho \rho'' + 2 \rho^2 \phi'^2 - 4 \alpha b D^2_3
\left[ (\omega \rho'^2 - 1) \phi' {\rm e}^{2 b \phi} \right]'
+ 2 \alpha D^2_3 {\rm e}^{2 b \phi} \left( 2 b \omega' \rho'^2 \phi'
- D^4_4 \frac{\omega \rho'^2 - 1}{\rho} \rho'' \right) = 0,
\label{Eqrho}
\end{equation}
\begin{eqnarray}
\rho \omega'' + 2 D^3_3 (\omega \rho')' + D^3_4 \frac{\omega \rho'^2 - 1}{\rho}
+ 2 \omega \rho \phi'^2 - 2 \rho f'^2 {\rm e}^{2 a \phi}
- 8 \alpha b D^3_3 \left( \omega \omega' \rho' \phi' {\rm e}^{2 b \phi} \right)' &&
\nonumber\\
- \alpha D^3_4 {\rm e}^{2 b \phi} \Biggl\{ D^5_6
\frac{(\omega \rho'^2 - 1)^2}{\rho^3} + 4 D^5_5 \left[ (\omega \rho')'
+ 2 b \omega \rho' \phi' \right] \frac{\omega \rho'^2 - 1}{\rho^2} &&
\nonumber\\
+ 2 \left[ \omega'' + 4b(\omega \phi')' + 8 b^2 \omega \phi'^2 \right]
\frac{\omega \rho'^2 - 1}{\rho} + 2 \frac{\rho'}{\rho}
\left[ 2 \omega \omega' \rho'' + \omega'^2 \rho' + 8 b \omega \phi'
(\omega \rho')' \right] \Biggr\} &=& 0.
\label{Eqw}
\end{eqnarray}

%%%%%%%%%%%%%%%%%%%%%%%%%%%%%%%%%%%%%%%%%%%%%%%%%%%%%%%%%%%%%%%%%%%%%%
\subsection{Symmetries of the reduced action}
\label{symmetries}
%%%%%%%%%%%%%%%%%%%%%%%%%%%%%%%%%%%%%%%%%%%%%%%%%%%%%%%%%%%%%%%%%%%%%%
One can check that equations of motion are invariant under a {\em
three-parametric} group of global transformations which consist of
the transformations of the field functions:
\begin{equation}
\label{symsol}
\omega \to \omega \, {\rm e}^{\mu}, \qquad \rho \to \rho \, {\rm
e}^{\delta}, \qquad \phi \to \phi + \frac{\delta}{b}, \qquad f \to f
\, {\rm e}^{\frac{\mu}2 - \frac{a}{b} \delta},
\end{equation}
accompanied by the  shift and rescaling of the radial variable
\begin{equation}
\label{trr}
r \to r \, {\rm e}^{\frac{\mu}2 + \delta} + \nu.
\end{equation}
Transformation of the electric potential is equivalent to rescaling
of the electric charge
\begin{equation}
q_e \to q_e \, {\rm e}^{\left( D - 3 + \frac{a}{b} \right) \delta}.
\end{equation}
Not all of these symmetries are the symmetries of the Lagrangian,
however. Integrating the action (\ref{action}) over the
$(D-2)$-dimensional sphere and dropping integration over time
integral, one obtains the one-dimensional reduced Lagrangian from
the relation $I = \int L dr$. Up to the total derivative one has:
\begin{eqnarray}
L &=& D^2_2 \rho' \left( \omega \rho^{D-3} \right)' + D^2_3 \rho^{D - 4}
-  2 \rho^{D - 2} (\omega \phi'^2 - f'^2 {\rm e}^{2 a \phi})
\nonumber\\
&-& \frac43 \alpha D^2_4 \rho'^3 \left( \omega^2 \rho^{D - 5} {\rm e}^{2 b \phi}
\right)' + 4 \alpha D^2_4 \rho' \left( \omega \rho^{D - 5} {\rm e}^{2 b \phi} \right)'
\nonumber\\
&-& \alpha {\rm e}^{2 b \phi} \left[ 4 b D^2_3 \rho^{D - 4} \omega' \phi'
- 2 D^2_4 \rho^{D - 5} \omega' \rho' - D^2_5 \rho^{D - 6} (\omega \rho'^2 - 1)
\right] (\omega \rho'^2 - 1).
\end{eqnarray}
It is easy to check that the one-dimensional action remains
invariant under the above transformations provided
\begin{equation}
\mu = - 2 (D - 3) \delta,
\end{equation}
namely under the following {\em two-parametric} group of global
transformations:
\begin{equation}
\label{symL}
r \to r \, {\rm e}^{- (D - 4) \delta} + \nu, \quad \omega \to \omega \,
{\rm e}^{- 2 (D - 3) \delta}, \quad \rho \to \rho \, {\rm e}^{\delta},
\quad \phi \to \phi + \frac{\delta}{b}, \quad f \to f \,
{\rm e}^{- \left( D - 3 + \frac{a}{b} \right) \delta}.
\end{equation}
They generate two conserved Noether currents
\begin{equation}
J_g := \left( \frac{\partial L}{\partial \Phi'^A} \Phi'^A - L \right)
\partial_g r \bigg|_{g=0} - \frac{\partial L}{\partial \Phi'^A} \,
\partial_g \Phi^A \bigg|_{g=0}, \qquad \partial_r J_g = 0,
\end{equation}
where $\Phi^A$ stands for $\omega, \rho, \phi, f$, and $g = \delta,
\nu$. The conserved quantity corresponding to $\nu$ is the
Hamiltonian
\begin{eqnarray}
\label{J1}
H &=& D^2_2 \rho' \left( \omega \rho^{D-3} \right)' - D^2_3 \rho^{D - 4}
- 2 \omega \rho^{D - 2} \phi'^2 + 2 \rho^{D - 2} f'^2 {\rm e}^{2 a \phi}
\nonumber\\
&-&\! \alpha {\rm e}^{2 b \phi} \left[ 4 b D^2_3 \rho^{D - 4} \omega'
\phi' (3 \omega \rho'^2 \!-\! 1) \!-\! 2 D^2_4 \rho^{D - 5} \omega'
\rho' (3 \omega \rho'^2 \!-\! 1) \!-\! D^2_5 \rho^{D - 6}
(\omega \rho'^2 \!-\! 1) (3 \omega \rho'^2 \!+\! 1) \right]
\nonumber\\
&-& 4 \alpha D^2_4 \rho'^3 \left( \omega^2 \rho^{D - 5} {\rm e}^{2 b \phi}
\right)' + 4 \alpha D^2_4 \rho' \left( \omega \rho^{D - 5}
{\rm e}^{2 b \phi} \right)'.
\end{eqnarray}
This is known to vanish on shell for diffeomorphism invariant theories,
$H = 0$. The Noether current corresponding to the parameter $\delta$
leads to the conservation equation $\partial_r J_\delta = 0$, where
\begin{eqnarray}
\label{J2}
J_\delta &=& - D^4_4 r H - D^2_2 \omega' \rho^{D-2} + \frac4{b}
\omega \rho^{D-2} \phi' + 4 \left( D - 3 + \frac{a}{b} \right) q_e f
\nonumber\\
&+& \alpha {\rm e}^{2 b \phi} \left[ (\omega \rho'^2 - 1)
( 2 D^2_2 D^2_3 \omega' \rho^{D-4} - 8 b D^2_3 \omega \rho^{D-4} \phi' )
+ 8 b D^2_3 \omega \omega' \rho^{D-3} \rho' \phi' \right].
\end{eqnarray}
Symmetry transformations will be used to rescale numerically obtained solutions to
desired asymptotic form and obtain true physical parameters of the
solution.

%%%%%%%%%%%%%%%%%%%%%%%%%%%%%%%%%%%%%%%%%%%%%%%%%%%%%%%%%%%%%%%%%%%%%%
%%%%%%%%%%%%%%%%%%%%%%%%%%%%%%%%%%%%%%%%%%%%%%%%%%%%%%%%%%%%%%%%%%%%%%
\section{Stretching the horizon of small black hole}
\label{stretch}
%%%%%%%%%%%%%%%%%%%%%%%%%%%%%%%%%%%%%%%%%%%%%%%%%%%%%%%%%%%%%%%%%%%%%%
%%%%%%%%%%%%%%%%%%%%%%%%%%%%%%%%%%%%%%%%%%%%%%%%%%%%%%%%%%%%%%%%%%%%%%

%%%%%%%%%%%%%%%%%%%%%%%%%%%%%%%%%%%%%%%%%%%%%%%%%%%%%%%%%%%%%%%%%%%%%%
\subsection{Small dilatonic $D$-dimensional black hole without GB term}
\label{small}
%%%%%%%%%%%%%%%%%%%%%%%%%%%%%%%%%%%%%%%%%%%%%%%%%%%%%%%%%%%%%%%%%%%%%%
Let us first discuss the black hole solution without the GB term. It
can be presented in the form~\cite{Gal'tsov:2005vf}
\begin{eqnarray}
&& ds^2 = - f_+ f_-^{-1 + \frac{4(D-3)}{(D-2)\Delta}} dt^2 + f_+^{-1} f_-^{-1
+ \frac2{D-3} - \frac4{(D-2)\Delta} } dr^2 + r^2 f_-^{\frac2{D-3}
- \frac4{(D-2)\Delta}} d\Omega_{D-2}^2,
\\
&& \mathrm{e}^{2 a \phi} = \mathrm{e}^{2 a \phi_\infty} f_-^{- \frac{2a^2}{\Delta}},
\qquad
F_{tr} = 4 (D - 3) \frac{(r_+ r_-)^\frac{D-3}{2}}{\sqrt{\Delta}}
\mathrm{e}^{- a \phi_\infty} \frac1{r^{D-2}},
\end{eqnarray}
where
\begin{equation}
f_\pm = 1 - \frac{r_\pm^{D-3}}{r^{D-3}}, \qquad \Delta = a^2 + \frac{2(D-3)}{D-2}.
\end{equation}
The mass and the electric and dilaton charges are given by
\begin{eqnarray}
\mathcal{M} &=& \frac{\Omega_{D-2}}{16 \pi G} \left[ (D-2) (r_+^{D-3}
- r_-^{D-3}) + \frac{4(D-3)}{\Delta} r_-^{D-3} \right],
\\
Q_e &=& \frac{(D-3) \Omega_{D-2}}{4 \pi G} \sqrt{\frac{(r_+ r_-)^{D-3}}{\Delta}} \;
\mathrm{e}^{a \phi_\infty},
\\
\mathcal{D} &=& - \frac{(D-3) a \Omega_{D-2}}{4 \pi G \Delta} r_-^{D-3}.
\end{eqnarray}

For $a = 0$, this solution reduces to the $D$-dimensional Reissner-Nordstr\"om
solution. In the extremal limit $r_+ = r_- = r_0$, it contracts to
\begin{equation}\label{ReNo}
ds^2 = - f_0^2 dt^2 + f_0^{-1} dr^2 + r^2 d\Omega_{D-2}^2, \qquad f_0
= 1 - \frac{r_0^{D-3}}{r^{D-3}},
\end{equation}
and has a degenerate event horizon $AdS_2 \times S^{D-2}$. Note that for
$a = 0$, the GB term decouples from the system, so this solution
remains valid in the full theory with $\alpha \neq 0$.

For $a \neq 0$, the extremal solution reads
\begin{equation}\label{edblh}
ds^2 = - f_0^{\frac{4(D-3)}{(D-2)\Delta}} dt^2 + f_0^{- \frac{2(D-4)}{D-3}
- \frac4{(D-2)\Delta} } dr^2 + r^2 f_0^{\frac2{D-3} - \frac4{(D-2)\Delta}} d\Omega_{D-2}^2.
\end{equation}
This has a null singularity at the horizon. The Ricci scalar in the
vicinity of this point diverges as
\begin{equation}
R \sim (r^{D-3} - r_0^{D-3})^{-\frac2{D-3} + \frac{4}{(D-2) \Delta}},
\end{equation}
together with the dilaton function
\begin{equation}
{\rm e}^{2 a \phi} \sim (r^{D-3} - r_0^{D-3})^{-\frac{2 a^2}{\Delta}}.
\end{equation}
The divergence of the GB term near the horizon is~\footnote{We use
this occasion to correct Eq.(33) of our previous paper for $D = 4$ \cite{Chen:2006ge}.}
\begin{equation}\label{GBdiv}
\mathrm{e}^{2 a \phi}{\cal L}_{GB}|_{r=r_+} \sim (r_+
- r_-)^{-\frac{a^2(D^2-4)}{a^2 (D-2) + 2 (D-3)}},
\end{equation}
so one can expect that the GB term will substantially modify the dilaton
black hole solution in the extremal limit.

The mass, the dilaton charge and the electric charge for this
solution (defined as in Sec.~IV below) are
\begin{equation}
\mathcal{M} = \frac{\Omega_{D-2}}{4 \pi G} \frac{D^3_3}{\Delta} \, r_0^{D-3}, \qquad
\mathcal{Q}_e = Q_e \mathrm{e}^{-a \phi_\infty}
= \frac{\Omega_{D-2}}{4 \pi G} \frac{D^3_3}{\sqrt{\Delta}} r_0^{D-3}, \qquad
\mathcal{D} = - \frac{\Omega_{D-2}}{4 \pi G} \frac{a D^3_3}{\Delta} r_0^{D-3}.
\end{equation}
They are determined by a single parameter $r_0$, so we have the
following relations among the three quantities
\begin{equation}
\mathcal{D} = a \mathcal{M}, \qquad \mathcal{Q}_e = \sqrt{\Delta} \mathcal{M},
\end{equation}
which imply the following BPS condition
\begin{equation}
a^2 \mathcal{M}^2 + \mathcal{D}^2 = \frac{2 a^2}{\Delta} \mathcal{Q}_e^2.
\end{equation}

%%%%%%%%%%%%%%%%%%%%%%%%%%%%%%%%%%%%%%%%%%%%%%%%%%%%%%%%%%%%%%%%%%%%%%
\subsection{Wiltshire black hole}
\label{wiltshire}
%%%%%%%%%%%%%%%%%%%%%%%%%%%%%%%%%%%%%%%%%%%%%%%%%%%%%%%%%%%%%%%%%%%%%%
Another limit in which our action admits an exact solution is
that of vanishing  dilaton. This is consistent with the field
equations for $a = b = 0$. In this case an exact solution was
found by Wiltshire~\cite{Wiltshire:1988uq}
\begin{equation}
\omega(r) = 1 + \frac{r^2}{2 D^3_4 \alpha} \left( 1 \mp \sqrt{1
+ \frac{64 \pi D^3_4 \alpha \mathcal{M}}{D^2_2 \Omega_{D-2} r^{D-1}}
- \frac{8 D^4_4 \alpha \, q_e^2}{D^2_2 r^{2(D-2)}}  } \right), \quad \rho(r) = r.
\end{equation}
The lower sign corresponds to an asymptotically AdS space-time for
$\alpha > 0$ and to an asymptotically de Sitter solution for $\alpha < 0$. The upper
sign leads to an asymptotically flat solution coinciding with the
$D$-dimensional Reissner-Nordstr\"om solution. These solutions exist
in dimensions $D \geq 5$ where the GB term is not the total
derivative. The asymptotically flat solution has two horizons which
coincide in the extremal limit for a special value of the electric
charge. For the extremal solution, the mass and the charge can be
expressed in terms of the single parameter, the radius of the horizon $r_0$:
\begin{equation}
\label{gpWe}
{\cal M} = \frac{\Omega_{D-2}}{8\pi} (D-2)[r_0^2+(D-4)^2 \alpha] r_0^{D-5}, \qquad
q_e^2 = \frac{D^2_3}{2} [r_0^2+ D^4_5 \alpha] r_0^{2(D-4)}.
\end{equation}
Conversely the radius can be expressed as
\begin{equation}
r_0^{D-3} = - \frac{4 \pi D^5_5 \mathcal{M}}{D^2_2 \Omega_{D-2}}
+ \sqrt{ \left( \frac{4 \pi D^5_5 \mathcal{M}}{D^2_2 \Omega_{D-2}} \right)^2
+ \frac{2 D_4^4 q_e^2}{D^2_3}}.
\end{equation}

%%%%%%%%%%%%%%%%%%%%%%%%%%%%%%%%%%%%%%%%%%%%%%%%%%%%%%%%%%%%%%%%%%%%%%
\subsection{Local solution near the horizon}
\label{local}
%%%%%%%%%%%%%%%%%%%%%%%%%%%%%%%%%%%%%%%%%%%%%%%%%%%%%%%%%%%%%%%%%%%%%%
In what follows we set $b = a$ as relevant for the heterotic string
theory case, but still keeping $a$ arbitrary. Assuming that the full
system with the GB term admits the $AdS_2 \times S^{D-2}$ horizon, $r
= r_H$, we look for the series expansions of the metric function in
powers of $x = r - r_H$:
\begin{equation}
\omega(r) = \sum_{i=2}^\infty \omega_i x^i, \qquad \rho(r) =
\sum_{i=0}^\infty \rho_i x^i, \qquad P(r) := {\rm e}^{2 a \phi(r)} =
\sum_{i=0}^\infty P_i x^i.
\end{equation}
The function $\omega$ starts with the quadratic term in view of the
degeneracy of the horizon, while two other functions have the
general Taylor's expansions. Denoting the physical radius of the
horizon $\rho_0 = \rho(r_H)$, we obtain for the leading order
coefficients:
\begin{eqnarray}
\label{NHSol}
\omega_2 = \frac{D^2_3}{2 \rho_0^2}, \qquad P_0 = \frac{\rho_0^2}{4 \alpha (2 D - 7)},
\end{eqnarray}
and $\rho_0$ is related to the electric charge via
\begin{equation}
\label{defqe}
\rho_0^{D-2} = q_e \frac{4 \sqrt{2 \alpha} (2D - 7)}{\sqrt{D^2_3
(D^2 - D - 8)}}.
\end{equation}
Note that the expression under the square root and the right hand as
a whole are positive for $D \geq 4$. The horizon radius is fixed
entirely by the electric charge, like in the extremal
Reissner-Nordstr\"om  case. In our units the GB parameter $\alpha$ has
dimension $L^2$. When the GB term is switched off ($\alpha \to 0$),
the horizon radius shrinks, as expected for small extremal black
holes. Higher order expansion coefficients exhibit dependence on
only one free parameter, namely the $P_1$ in the dilaton expansion.
Other coefficients are expressed in terms of the horizon radius
$\rho_0$ and $P_1$, the first sub-leading coefficients being
\begin{eqnarray}
\omega_3 &=& - \frac{2 \alpha P_1}{3 a^2 \rho_0^4 (D^3 - 9D^2 + 16D
+ 8)} \Bigl[ D^2_3 (3D^5 - 43D^4 + 213D^3 - 421D^2 + 236D + 76) a^2
\nonumber\\
&& + D^3_3 (2D - 7) (3D^4 - 25D^3 + 70D^2 - 56D - 40) \Bigr],
\nonumber\\
\rho_1 &=& \frac{4 \alpha P_1 [ (D^4 - 13D^3 + 54D^2 - 72D - 2) a^2
+ (2D - 7)(D^2 - 3D -2)]}{a^2 \rho_0 (D^3 - 9D^2 + 16D + 8)}.
\end{eqnarray}
%{\bf Some higher order coefficients are given in the Appendix.}
One can notice that the free parameter enters the expansion coefficients
always in the combination $P_1/a^2$. This facilitates transition to
the Wiltshire case.

In the limit of decoupled dilaton $a = 0$, the parameter $P_1 \to 0$,
while the ratio $P_1/a^2$ remains finite. In this case we have
nonvanishing coefficients $P_0, \; \rho_1$ and $\omega_i$:
\begin{equation}
P(r) = P_0, \qquad \rho(r) = \rho_0 + \rho_1 (r - r_0).
\end{equation}
The asymptotic flatness requires $\rho_1 = 1$, so we have $\rho_0 =
r_0$. For the Wiltshire solution $P_0 = 1$, and  we obtain the
following relation in the extremal case:
\begin{equation}
\rho_0^2 = r_0^2 = 4 \alpha (2D - 7),
\label{ddc}
\end{equation}
Substituting (\ref{ddc}) into (\ref{gpWe}), we find
\begin{equation}
\label{gpWe1}
{\cal M} = \frac{\Omega_{D-2}}{32\pi} \frac{D^2_2 (D^2 - 12)}{2D-7} \,
r_0^{D-3}, \qquad q_e^2 = \frac1{8} \frac{D^2_3 (D^2-D-8)}{2D-7}
\, r_0^{2(D-3)},
\end{equation}
and so our solution with the decoupled dilaton coincides with the extremal
case of the Wiltshire solution.

We can consider subgroup of global symmetry transformation
defined by two parameters $\delta$ and $\mu$.
We can eliminate the parameter $\rho_0$ from the expansion on the horizon
if we apply the transformation with parameters $\mu = -2 \ln\rho_0$
and $\delta = - \ln\rho_0$. The other transformation with parameters
$\mu = 2 \ln|P_1|, \, \delta = 0$ can take out $P_1$
from expansions. Choosing the absolute value $|P_1|$ in the second
transformation allows us to get the remaining parameter $\xi =\frac{P_1}{|P_1|}$
in the expansion at the horizon which fixes the sign of $\rho$
($\rho_1 > 0$ to obtain a global solution) in the expansion.
As a result, we have a map between parameters of expansion $P_1,\, \rho_0$ and
parameters of global transformation $\mu,\, \delta$. Typically one can first
investigate special solution with a simple choice of near horizon data, such as
$P_1 = 1$ and $\rho_0 = 1$, which are the values we use for our numerical analysis below.
Then general solutions with arbitrary values of free
parameters can be simply obtained by the global transformation with
$\mu = 2\ln\frac{|P_1|}{\rho_0}, \, \delta = -\ln\rho_0$.
Also it is clear from the relation between $\rho_0$ and $q_e$
that electrical charge plays the role of rescaling parameter. Finally the free
parameter $P_1$ could be fixed in accordance with the boundary condition at infinity.
We can also eliminate the GB coupling constant $\alpha$ from the system
by introducing new dilaton function $F = \alpha\, P$ and rescaling charge $q_e$.
In this way, we have only two parameters, the number of the dimension $D$ and
the dilaton coupling $a$, which affect the dynamics of solutions.

The values of the integrals of motion~(\ref{J1}) and (\ref{J2}) in terms of the
parameters of the local solution are
\begin{equation}
H = - \alpha \rho_0^{D-2} D^2_5 P_0 - D^2_3 \rho_0^{D-4}
+ \frac{2 q_e^2}{P_0} \rho_0^{2-D}, \qquad J_{\delta} = 4 (D-2) q_e f_0.
\end{equation}

%%%%%%%%%%%%%%%%%%%%%%%%%%%%%%%%%%%%%%%%%%%%%%%%%%%%%%%%%%%%%%%%%%%%%%
\subsection{The entropy}
\label{entropy}
%%%%%%%%%%%%%%%%%%%%%%%%%%%%%%%%%%%%%%%%%%%%%%%%%%%%%%%%%%%%%%%%%%%%%%
Knowledge of the local solution near the horizon is enough to
calculate the entropy of the black hole, assuming that the local
solution can be extended to infinity. To compute the entropy,
we apply Sen's entropy function approach~\cite{Sen:2007qy} which is
valid for the black holes with near horizon geometry of $AdS_2
\times S^{D-2}$. Using the notation of~\cite{Sen:2007qy} we
parametrize the near horizon geometry by two constants, $v_1$ and
$v_2$ related to the radii of $AdS_2$ and $S^{D-2}$, as
\begin{equation}
ds^2 = v_1 \left( - r^2 d\tau^2 + \frac{dr^2}{r^2} \right) + v_2
d\Omega^2_{D-2}.
\end{equation}
The scalar curvature and the GB term will read
\begin{equation}
R = - \frac2{v_1} + \frac{D^2_3}{v_2}, \qquad
\mathcal{L}_{GB} = \frac{D^2_5}{v_2^2} - \frac{4 D^2_3}{v_1 v_2}.
\end{equation}
The dilaton field and gauge field strength are constant on the
horizon
\begin{equation}
\phi = u, \qquad F_{\tau r} = p.
\end{equation}
Sen's entropy function is defined to be the integrand of the
action after integrating all angular coordinates of $S^{D-2}$. Using
(\ref{action}) we obtain
\begin{equation}
f = \frac{\Omega_{D-2}}{16 \pi G} v_1 v_2^{\frac{D-2}2} \left[ - \frac2{v_1}
+ \frac{D^2_3}{v_2} + \mathrm{e}^{2 a u}
\frac{2 p^2}{v_1^2} + \alpha \mathrm{e}^{2 a u} \left(
\frac{D^2_5}{v_2^2} - \frac{4 D^2_3}{v_1 v_2} \right) \right].
\end{equation}
The parameters $v_1, v_2, u, e$ are related to the near horizon
expansion coefficients by (note the rescaling of time coordinates
$\tau = \omega_2 t$)
\begin{equation}
\omega_2 = \frac1{v_1}, \qquad
\rho_0 = \sqrt{v_2}, \qquad
P_0 = \mathrm{e}^{2 a u}, \qquad
q_e = p \, v_1^{-1} \, v_2^{\frac{D-2}2} \,\mathrm{e}^{2 a u}.
\end{equation}

According to the equations of motion, the value of parameters should
minimize the entropy function:
\begin{equation}
\partial_{v_1} f = 0, \qquad
\partial_{v_2} f = 0, \qquad
\partial_u f = 0,
\end{equation}
which lead to the following constraints
\begin{equation}
v_2 = \frac{D^2_3}2 v_1, \qquad
v_1 = \frac{4 (2 D - 7) p^2}{D^2 - D- 8} \mathrm{e}^{2 a u}, \qquad
p^2 = \frac{2(D^2 - D - 8)}{D^2_3}
\alpha,
\end{equation}
and furthermore imply $f = 0$. These three constraints are exactly
identical with the relations (\ref{NHSol}) and (\ref{defqe}) from the
near horizon analysis. The physical electric charge, $q$ (i.e. $Q_e$ defined
in subsection~\ref{global}), can be
obtained via $q = \partial_e f$
\begin{equation}
q = \frac{\Omega_{D-2}}{4 \pi G} \; p \, v_1^{-1} \,
v_2^{\frac{D-2}2} \mathrm{e}^{2 a u} =
\frac{\Omega_{D-2}}{4 \pi G} \; q_e.
\end{equation}
The entropy of black holes is related to the entropy function by a
Legendre transformation
\begin{equation}
S = 2 \pi (q p - f) = 2 \pi q p = \frac{D^2 - D - 8}{8 (2 D - 7) G} \;
\Omega_{D-2} v_2^{\frac{D-2}2}.
\end{equation}

The horizon area of $AdS_2 \times S^2$ is
$A = \mathrm{vol}(\Omega_{D-2}) v_2^{\frac{D-2}2}$, thus the entropy can be
expressed in terms of area of horizon as
\begin{equation}
S = \frac{D^2 - D - 8}{8 (2 D - 7) G} A = \frac{A}{4 G} + \frac{D^2 - 5
D + 6}{8 (2 D - 7) G} A  = S_{BH} + S_{GB},
\end{equation}
and the deviation of the entropy from Bekenstein-Hawking relation
by the GB term increases for higher and higher dimensions. For
example, the ratio of $S_{GB}/S_{BH}$ from $D = 4$ to $10$ is
\begin{equation}
\frac{S_{GB}}{S_{BH}} = \left\{ 1, 1, \frac65, \frac{10}7, \frac53,
\frac{21}{11}, \frac{28}{13} \right\}.
\end{equation}
A general discussion on the entropy of theories with quadratic
curvature correction and Lovelock theory is given in
\cite{Cai:2007cz}.

%%%%%%%%%%%%%%%%%%%%%%%%%%%%%%%%%%%%%%%%%%%%%%%%%%%%%%%%%%%%%%%%%%%%%%
%%%%%%%%%%%%%%%%%%%%%%%%%%%%%%%%%%%%%%%%%%%%%%%%%%%%%%%%%%%%%%%%%%%%%%
\section{Asymptotics}
\label{asymptotics}
%%%%%%%%%%%%%%%%%%%%%%%%%%%%%%%%%%%%%%%%%%%%%%%%%%%%%%%%%%%%%%%%%%%%%%
%%%%%%%%%%%%%%%%%%%%%%%%%%%%%%%%%%%%%%%%%%%%%%%%%%%%%%%%%%%%%%%%%%%%%%
Now consider the asymptotic expansions of the metric function
by substituting the following expansions into the equations of motion:
\begin{equation}
\omega(r) = 1 + \sum_{i=1} \frac{\bar\omega_i}{r^i}, \qquad \rho(r)
= r + \sum_{i=1} \frac{\bar\rho_i}{r^i}, \qquad \phi(r) =
\bar\phi_\infty + \sum_{i=1} \frac{\bar\phi_i}{r^i}
\end{equation}
According to the falloff of the Newton potential in different
dimensions, one has the first non-zero term in the expansion for
$\omega$ and that for dilaton starting from $i = D-3$, while $\rho$
differs from $r$ in $(2D-7)$-th terms:
\begin{eqnarray}
\omega(r) &=& 1 + \frac{\bar\omega_{D-3}}{r^{D-3}} + \frac{2 q_e^2
\, \mathrm{e}^{- 2 a \bar\phi_\infty}}{D^2_3} \frac1{r^{2(D-3)}} +
O\left( \frac1{r^{2D-4}} \right),
\nonumber\\
\rho(r) &=& r - \frac{(D-3) \bar\phi_{D-3}^2}{(D - 2)(2D - 7)}
\frac1{r^{2D - 7}} + O\left( \frac1{r^{2D-5}} \right),
\\
\phi(r) &=& \bar\phi_\infty + \frac{\bar\phi_{D-3}}{r^{D-3}} -
\frac12 \left[ \frac{a q_e^2 \, \mathrm{e}^{- 2 a
\bar\phi_\infty}}{(D-3)^2} + \bar\omega_{D-3} \bar\phi_{D-3} \right]
\frac1{r^{2(D-3)}} + O\left( \frac1{r^{2D-4}} \right). \nonumber
\end{eqnarray}
One can notice, that these terms of expansion do not contain the
GB coupling $\alpha$. The contribution of the GB term is manifest in
the third non-vanishing coefficient in $\rho$.
If the GB term is switched off $\alpha = 0$, the third
non-vanishing coefficient of $\rho$  is
\begin{equation}
\bar\rho_{3D-10} = \frac{4}{3 D^2_3 (3D-10)} \bar\phi_{D-3} \left[
(D-3)^2 \bar\omega_{D-3} \bar\phi_{D-3} + a q_e^2 \, \mathrm{e}^{- 2
a \bar\phi_\infty} \right].
\end{equation}
In presence of the GB term it is
\begin{equation}
\bar\rho_{2D-5} = \frac{2(D-3)^2}{2D-5} \alpha a \bar\omega_{D-3}
\bar\phi_{D-3} \mathrm{e}^{2 a \bar\phi_\infty}.
\end{equation}
So in $D = 4$ the GR contribution dominates appearing as
$\bar\rho_2$, and in $D = 5$ both GR and GB contributions appear in
$\bar\rho_5$, but in higher dimensions, GB contribution is leading.

For the asymptotically flat geometry the global physical quantities,
such as mass and charges, can be read out from the asymptotic
expansion. Since the first sub-leading coefficients are independent of
the GB coupling, we can still use the formula of global charges for
the theories without higher curvature corrections.

%%%%%%%%%%%%%%%%%%%%%%%%%%%%%%%%%%%%%%%%%%%%%%%%%%%%%%%%%%%%%%%%%%%%%%
\subsection{Global charges}
\label{global}
%%%%%%%%%%%%%%%%%%%%%%%%%%%%%%%%%%%%%%%%%%%%%%%%%%%%%%%%%%%%%%%%%%%%%%
The ADM mass is given in our notation by~\footnote{The volume of
$S^{D-2}$ is $\Omega_{D-2} = \frac{2 \pi^{\frac{D-1}2}}{\Gamma(\frac{D-1}2)}$
and the gamma function is either $\Gamma(n+1) = n!$ or $\Gamma(\frac{n}2+1) = \sqrt\pi
\frac{n!!}{2^{\frac{n+1}2}}$ for integer $n$. This gives
$\Omega_{D-2} = \left\{ 4\pi, 2\pi^2, \frac83 \pi^2,
\pi^3, \frac{16}{15} \pi^3, \frac13 \pi^4, \frac{32}{105} \pi^4
\right\}$ for $D = 4, \cdots, 10$.}
\begin{equation}
\mathcal{M} = \frac{\Omega_{D-2}}{8 \pi G} (D - 2)
\left[ r^{D-3} \left( \frac1{\sqrt\omega} - \frac{\rho}{r} \right)
- r^{D-2} \left( \frac{\rho}{r} \right)' \right]_{r \to \infty},
\end{equation}
and reduces to
\begin{equation}
\mathcal{M} = - \frac{\Omega_{D-2}}{16 \pi G} (D - 2)
\left[ \bar\omega_{D-3} - 2 (D-4) \bar\rho_{D-4} \right].
\end{equation}
For $D > 4$, in general, the ADM mass could depend not only on the
first sub-leading coefficient $\bar\omega_{D-3}$ of $\omega$, but
also on the sub-leading coefficient $\bar\rho_{D-4}$ of $\rho$.
But we have seen this coefficient is zero, so we have
\begin{equation}
\bar\omega_{D-3} = - \frac{16 \pi G \mathcal{M}}{(D-2) \Omega_{D-2}}.
\end{equation}

The definition of the dilaton charge $\mathcal{D}$ is
\begin{equation}
\mathcal{D} = \frac1{4 \pi G} \int_{r \to \infty} d\Omega_{D-2} \,
r^{D-2} \, \partial_r \phi,
\end{equation}
which has a contribution from the expansion coefficient
\begin{equation} \bar\phi_{D-3} = -\frac{4 \pi G \mathcal{D}}{(D-3) \Omega_{D-2}}.
\end{equation}

The physical electric charge can be computed by the flux
\begin{equation}
Q_e = \frac1{4 \pi G} \int_{r \to \infty} d\Omega_{D-2} \, r^{D-2} \,
\mathrm{e}^{2 a \phi_\infty} \, F_{tr},
\end{equation}
so we have the following relation between this quantity and the
charge introduced as an integration constant in the previous section:
\begin{equation}
q_e = \frac{4 \pi G}{\Omega_{D-2}}  Q_e.
\end{equation}

The asymptotic values of two integrals of motion~(\ref{J1}) and (\ref{J2}) are
\begin{eqnarray}
H^\infty &=& \left[ D_3^2 (r \rho')_\infty^{D-4}
- \alpha D_5^2 P_{\infty}(r \rho')_\infty^{D-6} \right] (\omega_\infty
\rho'_\infty{}^2 - 1),
\label{J1a}
\\
J_\delta^\infty &=& \left[ \alpha \mathrm{e}^{2 a \phi_\infty} D^2_5 D^4_4
 (r \rho')_\infty^{D-5} (\omega_\infty \rho_\infty'^2 - 1)
- D^2_4 (r \rho')_\infty^{D-3} \right] \frac{\omega_\infty \rho_\infty'^2
- 1}{\rho_\infty'}
\nonumber\\
&& + 4 D_2^2 q_e f_\infty - 2 D_2^2 D_3^2 \mathcal{M} \rho_\infty'^{D-2}
- 4 D_3^3 \omega_\infty \rho_\infty'^{D-2} \frac{\mathcal{D}}{a}.
\end{eqnarray}
From $H = 0$, we can see that $\omega_\infty \rho_\infty'^2 \to 1$ for
$r \to \infty$, which also regularizes the second integral of motion.
For Minkowski space ($\omega_\infty = \rho'_{\infty} = 1$), we have
\begin{equation}
J_\delta^\infty = 4 D_2^2 q_e f_\infty - 2 D_2^2 D_3^2 \mathcal{M}
- 4 D_3^3 \frac{\mathcal{D}}{a}.
\end{equation}
It is possible to apply global transformation to satisfy asymptotically flat
condition which fixes one of the free parameters in expansion around horizon,
$P_1$. Note that the values of the integral of motion are four times of what
we have in~\cite{Chen:2006ge}:
\begin{equation*}
H = \frac12 \left( \omega_\infty \rho'^2_\infty - 1 \right), \quad
J_\delta = 2 \, q_e \,f_{\infty} - \mathcal{M} - \frac{\mathcal{D}}a.
\end{equation*}

%%%%%%%%%%%%%%%%%%%%%%%%%%%%%%%%%%%%%%%%%%%%%%%%%%%%%%%%%%%%%%%%%%%%%%
\subsection{BPS condition}
\label{bpsCond}
%%%%%%%%%%%%%%%%%%%%%%%%%%%%%%%%%%%%%%%%%%%%%%%%%%%%%%%%%%%%%%%%%%%%%%
The theory we are considering here does not necessarily have an underlying
supersymmetry. However, it is instructive to investigate the fulfilment
of the no-force condition which is usually associated with the supersymmetry.
In particular, in the $D = 4$ case the supersummetric embedding into the heterotic
string theory in the supergravity limit gives the BPS condition for the extremal
small black holes ($\mathcal{Q}_e = Q_e \, \mathrm{e}^{- a \phi_\infty}$)
\begin{equation}
\mathcal{M}^2 + \mathcal{D}^2 = \mathcal{Q}_e^2.
\end{equation}
This corresponds to vanishing of the sum of the gravitational and dilaton
attractive forces and the electric repulsion. This does not hold if the
GB term is turned on. However, it was demonstrated in~\cite{Chen:2006ge}
that on the boundary of the allowed domain of the
dilaton coupling the role of the GB term is diminished, and the BPS
condition is restored.
Our aim here is to confirm this property.

In the higher-dimensional cases the gravitational, Coulomb and dilaton forces are
\begin{eqnarray}
F_g &\sim& - \frac{8 \pi G (D-3)}{(D-2) \Omega_{D-2}} \frac{\mathcal{M}^2}{r^{D-2}},
\nonumber\\
F_A &\sim& \frac{4 \pi G}{\Omega_{D-2}} \frac{\mathcal{Q}_e^2}{r^{D-2}},
\nonumber\\
F_\phi &\sim& - \frac{4 \pi G}{\Omega_{D-2}} \frac{\mathcal{D}^2}{r^{D-2}},
\end{eqnarray}
so the no force condition reads
\begin{equation}
\label{noforce}
2 (D - 3) \mathcal{M}^2 + (D - 2) \mathcal{D}^2 = (D - 2)
\mathcal{Q}_e^2.
\end{equation}
In the case that the GB term is decoupled, i.e. $\alpha = 0$,
the no-force condition (\ref{noforce}) at infinity is equivalent to the
degenerated horizon obtained for the exact extremal
dilatonic black hole solutions in the previous section
\begin{equation}
\mathcal{D} = a \mathcal{M}, \quad
\mathcal{Q}_e = \sqrt{\Delta} \mathcal{M} \quad
\Rightarrow \quad a^2 \mathcal{M}^2 + \mathcal{D}^2
= \frac{2 a^2}{\Delta} \mathcal{Q}_e^2.
\label{bps}
\end{equation}
Note that the relation (\ref{noforce}) does not involve explicitly
the dilaton coupling (though it appears in the definition of
$\mathcal{Q}_e$). The special case of $D = 4$ ($\Delta = a^2 + 1$) was earlier
discussed in~\cite{Chen:2006ge} in which case $\bar\omega_1 = - 2 \mathcal{M},
\bar\phi_1 = - \mathcal{D}$ (with different sign convention).

%%%%%%%%%%%%%%%%%%%%%%%%%%%%%%%%%%%%%%%%%%%%%%%%%%%%%%%%%%%%%%%%%%%%%%
%%%%%%%%%%%%%%%%%%%%%%%%%%%%%%%%%%%%%%%%%%%%%%%%%%%%%%%%%%%%%%%%%%%%%%
\section{Cusps}
\label{cusp}
%%%%%%%%%%%%%%%%%%%%%%%%%%%%%%%%%%%%%%%%%%%%%%%%%%%%%%%%%%%%%%%%%%%%%%
%%%%%%%%%%%%%%%%%%%%%%%%%%%%%%%%%%%%%%%%%%%%%%%%%%%%%%%%%%%%%%%%%%%%%%
It was discovered in~\cite{Chen:2006ge} that the 4D EGBD static
spherically symmetric gravity typically develops cusps at some
points $r = r_c$ in the vicinity of where the metric functions vanish.
There they have Taylor expansions in terms of
\begin{equation}
y = |r - r_c|.
\end{equation}
The metric and its first derivative are regular there, while the second
derivatives diverge as $y^{-1/2}$. There are therefore the cusp hypersurfaces
which are the spheres $S^{D-2}$ of finite radius. These cusp spheres
have curvature singularity which is rather mild (the Ricci scalar
diverges only as $y^{-1/2}$, and the Kretchmann scalar as $1/y$).
They are in fact the singular turning points of the radial variable $\rho(r)$.

The presence of turning points in the numerical solutions was encountered in
the case $D = 4$ in~\cite{Alexeev:1996vs, Pomazanov:2000, Chen:2006ge}.
The numerical solution can be extended through these points using the
technique of~\cite{Pomazanov:2000} and then the solution evolves into a strong
singularity. Here we find that the situation is similar in
higher dimensions $D \leq 6$, but starting from $D = 7$ the solution
can be extended to an asymptotically flat one.

%%%%%%%%%%%%%%%%%%%%%%%%%%%%%%%%%%%%%%%%%%%%%%%%%%%%%%%%%%%%%%%%%%%%%%
\subsection{Expansion near the turning points}
\label{expansion}
%%%%%%%%%%%%%%%%%%%%%%%%%%%%%%%%%%%%%%%%%%%%%%%%%%%%%%%%%%%%%%%%%%%%%%
The general property of the  turning points is that the metric
functions and the exponential of the dilaton field, $f = \{
\omega(r), \rho(r), F(r) = \alpha \mathrm{e}^{2 a \Phi(r)} \}$, have
finite first derivative and divergent second derivative, i.e.
\begin{equation}
f'(r_\mathrm{tp}) = \mathrm{constant}, \qquad f''(r_\mathrm{tp}) \to \infty.
\end{equation}
The metric functions and the dilaton can be expanded  in terms of
the fractional powers of the variable $y$
\begin{equation}
f(y) = f_0 + \sum\limits_{i=2} f_i \, y^{\frac{i}2},
\end{equation}
where we have either $y = r_\mathrm{tp} - r$ (the right turning point)
or $y = r - r_\mathrm{tp}$ (the left turning point). These two types
of turning points have opposite signs of the odd-order derivatives,
\begin{equation}
f^{(2n+1)}(r - r_\mathrm{tp}) = - f^{(2n+1)}(r_\mathrm{tp} - r),
\end{equation}
and the expansion coefficients, $\{ f_i \}$, have the same ``iterative'' relations for both
type turning points. The expansions read
\begin{eqnarray}
\omega &=& \omega_0 + \omega_2 \, y + \omega_3  \, y^{\frac32} +
O(y^2),
\\
\rho &=& \rho_0 + \rho_2 \, y + \rho_3  \, y^{\frac32} + O(y^2),
\\
F &=& F_0 + F_2  \, y + F_3 \, y^{\frac32} + O(y^2).
\end{eqnarray}
They contain four free parameters, namely $\omega_0,\, \rho_0,\,
F_0$ and $\rho_2$ (for the fixed charge parameter $q_e$), other
coefficients depending on them. The coefficient $\rho_3$ is given by
the square roots of a second order equation which can have two
branches (positive and negative) corresponding to double valued
solution near turning points. Similarly, $\omega_3$ and $F_3$ also
have two-branch solutions.

The exponents in the turning point expansions are independent of
$D$. Therefore, the rate of divergence of geometric quantities is
universal. More precisely, the scalar curvature is
\begin{equation}
R \sim - \frac34 \frac{\omega_3 \rho_0 + 2(D-2) \omega_0 \rho_3}{\rho_0} y^{-1/2},
\end{equation}
and the matter stress tensor is finite. Indeed, $\omega_3$ is
proportional to $\rho_3$ which has double values (with opposite
sign) near the turning point. Therefore, the sign of divergent
scalar curvature also changes. Moreover, one expects that the GB
combination should have $y^{-1}$ divergence, but actually it is
weaker, namely $y^{-1/2}$.

For numerical integration, we rewrite the equations of motion as a
matrix equation of the dynamical system
\begin{equation}
\bm{A} \bm{x}' = \bm{b},
\end{equation}
where 6D vector $\bm{x}$ denotes $\bm{x}(r) = \{ \omega(r),
\omega'(r), \rho(r), \rho'(r), F(r), F'(r) \}$.
The solution is ill-defined at the points where $\det\bm{A} = 0$.
The turning points are special cases of the general situation
(see~\cite{Pomazanov:2000} for complete classification). We can extend
solutions through the turning points  introducing a suitable new
parameter $\sigma$:
\begin{equation}
\dot r = \frac{dr}{d\sigma} = \lambda \det\bm{A},
\end{equation}
and generalizing the dynamical system to one dimension more, i.e.
$\tilde{\bm{x}}(\sigma) = \{ \omega(\sigma), \dot\omega(\sigma),
\rho(\sigma), \dot\rho(\sigma), F(\sigma), \dot F(\sigma), r(\sigma) \}$.
The matrix equation then becomes
\begin{equation}
\tilde{\bm{A}} \dot{\tilde{\bm{x}}} = \tilde{\bm{b}},
\qquad \tilde{\bm{A}} = \left( \begin{array}{cc} \bm{A} & -
\bm{b} \\ 0 & 1 \end{array} \right), \qquad \tilde{\bm{b}} =
\left( \begin{array}{c} 0 \\ \lambda \det\bm{A} \end{array} \right).
\end{equation}
The parameter $\lambda$ can be fixed by normalization of
$\dot{\tilde{\bm{x}}}$. We choose $\dot{\tilde{\bm{x}}}^2 =
\dot{\bm{x}}^2 + \dot r^2 = 1$ and this ensures that
$\dot r$ is finite for both small and large values of $\det\bm{A}$
which is useful for numerical calculation.

In terms of $\sigma$, we have the following result near turning point
\begin{equation}
r(\sigma) = r_\mathrm{tp} + r_1 (\sigma_\mathrm{tp} - \sigma)^2 + \cdots,
\end{equation}
where $r_1 < 0$ for  the right turning points and $r_1 >0$ for the
left ones. The metric can be rewritten as
\begin{equation}
\label{metricsigma}
ds^2 = - \omega(\sigma) dt^2 + \frac{d^2\sigma}{W(\sigma)}
+ \rho^2(\sigma) d\Omega_{D-2}^2,
\end{equation}
where $W = \omega / \dot r^2$. Now, the functions $\omega, W, \rho$
are single valued functions of $\sigma$.

%%%%%%%%%%%%%%%%%%%%%%%%%%%%%%%%%%%%%%%%%%%%%%%%%%%%%%%%%%%%%%%%%%%%%%
\subsection{Geodesics near the turning points}
\label{geodesics}
%%%%%%%%%%%%%%%%%%%%%%%%%%%%%%%%%%%%%%%%%%%%%%%%%%%%%%%%%%%%%%%%%%%%%%
As mentioned before, the curvature weakly diverges at turning point. One
would expect the property near turning points is much better than near
the singularity. So let us check the radial geodesic of $t$ and $\sigma$ as
functions of the proper time $\lambda$ ($d\theta/d\lambda = 0 =
d\phi/d\lambda$). The relevant Christoffel symbols are
\begin{equation}
\Gamma^t{}_{t\sigma} = \frac12 \frac{\dot \omega}{\omega}, \qquad \Gamma^\sigma{}_{tt}
= \frac12 W \dot \omega, \qquad \Gamma^\sigma{}_{\sigma\sigma}
= - \frac12 \frac{\dot W}{W},
\end{equation}
The geodesic equation for $t$
\begin{equation}
\label{GeoT}
\frac{d^2 t}{d\lambda^2} + \frac{\dot \omega}{\omega} \frac{dt}{d\lambda}
\frac{d\sigma}{d\lambda} = 0,
\end{equation}
can be simplified as
\begin{equation}
\frac{d}{d\lambda}\left( \omega \frac{dt}{d\lambda} \right) = 0,
\end{equation}
or
\begin{equation}
\frac{dt}{d\lambda} = \frac{C}{\omega},
\end{equation}
where the integration constant $C > 0$ means the ``energy'' per unit
mass of test particle at infinity. The geodesic equation for $\sigma$
coordinate is
\begin{equation}
\label{Geqsigma}
\frac{d^2 \sigma}{d\lambda^2} + \frac12 \left[ W \dot \omega \left(
\frac{dt}{d\lambda} \right)^2 - \frac{\dot W}{W} \left( \frac{d\sigma}{d\lambda}
\right)^2 \right] = 0.
\end{equation}
After integration, it reduces to
\begin{equation}
k = \omega \left( \frac{dt}{d\lambda} \right)^2 - \frac1{W} \left(
\frac{d\sigma}{d\lambda} \right)^2,
\end{equation}
or
\begin{equation}
\left(\frac{d\sigma}{d\lambda} \right)^2 = W \left( \frac{C^2}{\omega} - k \right)
= \frac{C^2 - k \omega}{\dot r^2} ,
\end{equation}
where $k = 1$ for time-like geodesic and $k = 0$ for null geodesic.

The geodesic solutions are
\begin{equation}
\label{solsigma2}
t = \frac{C}{\omega_0} \lambda, \qquad (\sigma - \sigma_\mathrm{tp})^2
= - \frac{k \omega_2}{4 r_1^2} \lambda^2 \pm \frac{\sqrt{C^2 - k \omega_0}}{|r_1|} \lambda.
\end{equation}
There are two possible solutions for $\sigma$, the minus branch is valid
for $-\infty < \lambda \le 0$ and plus branch for $0 \le \lambda < \infty$
and both geodesics are terminated at the turning point ($\lambda = 0$).
The only possible extension for the geodesic solution is ``gluing''
these two solutions, i.e.
\begin{equation}
(\sigma - \sigma_\mathrm{tp})^2 = - \frac{k \omega_2}{4 r_1^2} \lambda^2
+ \mathrm{sign}(\lambda) \frac{\sqrt{C^2 - k \omega_0}}{|r_1|} \lambda.
\end{equation}
However, one can easy check that the result is not a smooth solution at
$\lambda = 0$ and the second derivative of $\sigma$ with respect to $\lambda$
generates a delta function. Therefore, such extension, in general, is not
a solution of (\ref{Geqsigma}). Similar situation happens for the time-like
geodesic. Hence, the cusp turning points are not extendable for the geodesics.
However there is an exception for the special values of $C^2 = k \omega_0$ and
$k \omega_2 < 0$ (if $\omega_2 > 0$ the time-like geodesic cannot reach
the turning point).

It is instructive to compare divergences in various cases.
To study this, we use Krechman scalar $K =
R^{\alpha\beta\gamma\delta} R_{\alpha\beta\gamma\delta}$.
The Schwarzschild black hole at $r = 0$ has $K \sim r^{-2(D-1)}$,
and the Reissner-Nordstr\"{o}m black hole at $r = 0$ has $K \sim r^{-4(D-2)}$,
GB pure gravity black hole ($D > 5$) at $r=0$ has 1) $K \sim r^{-2(D-1)}$
for charged solution, and 2) neutral one $K \sim r^{-(D-1)}$.
The extremal dilatonic black hole with GB term at a turning point
$K \sim y^{-1}$ for any dimension:
\begin{equation}
R = \pm \frac3{4\rho_0} \frac{2 \, w_0 \rho_3 (D-2) +
\rho_0\, w_3}{\sqrt{y}},
\end{equation}
where $y = r_{tp} - r$ for upper sign (a first type of turning point)
and $y = r - r_{tp}$ for lower (a second type).
If $\rho_3 = 0$, the scalar curvature at such a type of point is regular,
but it imposes some constraint on the parameters
because $\rho_3$ depends on the parameters.

%%%%%%%%%%%%%%%%%%%%%%%%%%%%%%%%%%%%%%%%%%%%%%%%%%%%%%%%%%%%%%%%%%%%%%
%%%%%%%%%%%%%%%%%%%%%%%%%%%%%%%%%%%%%%%%%%%%%%%%%%%%%%%%%%%%%%%%%%%%%%
\section{Numerical Results}
\label{numerical}
%%%%%%%%%%%%%%%%%%%%%%%%%%%%%%%%%%%%%%%%%%%%%%%%%%%%%%%%%%%%%%%%%%%%%%
%%%%%%%%%%%%%%%%%%%%%%%%%%%%%%%%%%%%%%%%%%%%%%%%%%%%%%%%%%%%%%%%%%%%%%
Lets consider the special limit $a \to 0$ in which the dilaton field decouples.
The analytical general solution is
\begin{equation}
\omega(r) = 1 + \frac{r^2}{2 D^3_4 \alpha} \left( 1 \mp \sqrt{1 +
\frac{64 \pi D^3_4 \alpha \mathcal{M}}{(D-2) \mathrm{vol}(\Omega_{D-2})
r^{D-1}} - \frac{8 D^4_4 \alpha \, q_e^2}{D^2_2 r^{2(D-2)}} +
\frac{4 D^3_4 \alpha \Lambda}{D^1_2}} \right), \quad \rho(r) = r.
\end{equation}
For the case $\Lambda = 0$, the radius of degenerated (extremal)
horizon is $r_0^2 = 4 (2D - 7) \alpha$ which can be obtained from (\ref{ddc}).
It is clear that the horizon shrinks to a point when we turn off the GB term.
Although the dilaton field is decoupled, in the dimensions $D \geq 5$, the
GB term still gives a non-trivial contribution to
equations of motion which will break the BPS (non-force) condition.
In more detail, the ratio of the mass and the electrical charge for
an extremal solution can be computed
\begin{equation}
\frac{\Delta \, \mathcal{M}}{Q_e^2} = \frac{(D^2 - 12)^2}{4 (D^2 - D - 8)
(2D - 7)} \ge 1,
\end{equation}
and our numerical analysis give consistent results in the decoupling limit.
Moreover, the numerical analysis also indicates that the dilaton charge is
proportional to the dilation coupling times mass and
\begin{equation}
\frac{a^2 \mathcal{M}^2}{\mathcal{D}^2} \ge 1,
\end{equation}
but the exact form for the ratio is still unknown. In these two ratios,
the equality holds for $D = 4$ which saturates the BSP condition~(\ref{bps}).

In the numerical results, we are going to show the following quantities. From the
symmetries (\ref{symL}), we know that if we regard the electrical charge as a scaling
parameter, the following ratios will depend only on the dilaton coupling:
\begin{equation}
k_M(a) = \frac{\mathcal{M}^\frac{D-2}{D-3}}{Q_e}, \qquad
k_D(a) = \frac{\mathcal{D}^\frac{D-2}{D-3}}{Q_e}, \qquad
k_F(a) = \frac{\alpha \, \mathrm{e}^{(D-2) a \phi_\infty}}{Q_e}.
\label{ratio}
\end{equation}
For verifying the BPS conditions, we will analyze the ratios
\begin{equation}
\frac{a^2 \mathcal{M}^2}{\mathcal{D}^2}, \qquad \frac{\Delta \,
\mathcal{M}^2}{\mathcal{Q}_e^2}, \qquad k_{BPS}
= \frac{\Delta (a^2 \mathcal{M}^2 + \mathcal{D}^2)}{2 a^2 \mathcal{Q}_e^2}.
\end{equation}

We now present our numerical results.

%%%%%%%%%%%%%%%%%%%%%%%%%%%%%%%%%%%%%%%%%%%%%%%%%%%%%%%%%%%%%%%%%%%%%%
\subsection{$D = 4$ }
\label{d4}
%%%%%%%%%%%%%%%%%%%%%%%%%%%%%%%%%%%%%%%%%%%%%%%%%%%%%%%%%%%%%%%%%%%%%%
For convenience, we first recall the results for $D = 4$ found
in~\cite{Chen:2006ge}. Starting with small $a$ (for $a=0$ one has the
Reissner-Nordstr\"om solution) one finds that asymptotically flat
black holes with degenerate event horizons exist up to $a = a_{\rm cr}
= 0.488219703$. The critical value of dilaton coupling $a_{\rm cr}$
separates the regions where there exist regular asymptotically flat
solutions for $a < a_{\rm cr}$ and the singular ones for $a > a_{\rm cr}$,
which firstly have a throat ($\rho' = 0$), then a turning point where
$\rho''$ changes sign and $\rho''$ diverges and finally a singular point.
In the limit $a \to a_{\rm cr}$, the mass $\mathcal{M}$
diverges, and somewhat surprisingly, the BPS condition of the
theory without curvature corrections holds for the ratios of
parameters. This can be understood as dominance of the Einstein
term over Gauss-Bonnet. Indeed, if we keep the mass
fixed, the limit corresponds to gravitational constant $G$ going
to zero. Then the Einstein term becomes greater, unless the GB
term is increasing similarly.

The critical value of the dilaton coupling is less than the
heterotic string value $a=1$ or $1/2$, so no asymptotically flat extremal
EMDGB black holes exist in $D=4$.

%%%%%%%%%%%%%%%%%%%%%%%%%%%%%%%%%%%%%%%%%%%%%%%%%%%%%%%%%%%%%%%%%%%%%%
\subsection{$D = 5$ and $6$}
\label{d5}
\begin{figure}[ht]
\includegraphics[width=12cm]{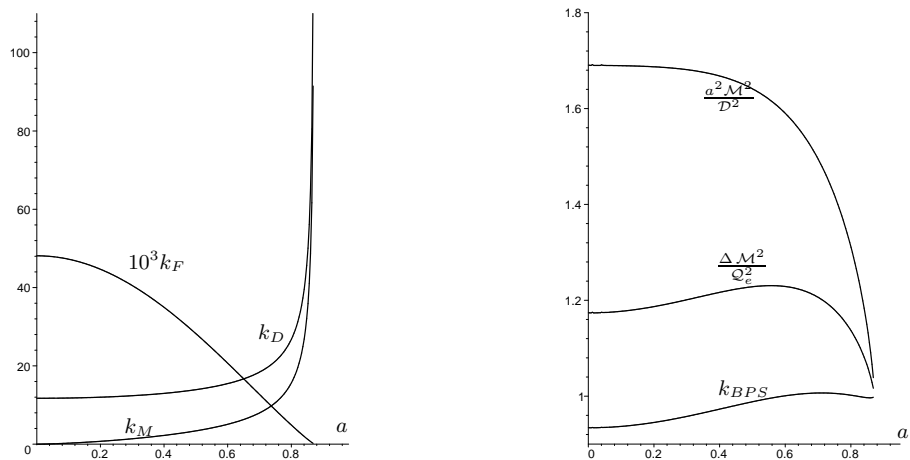}
\caption{Left: %ratios of mass, dilaton charge and the exponential
%of dilaton at spatial infinity to the electrical charge $Q_e$, i.e.
$k_M(a), k_D(a)$ and $k_F(a)$ (multiplied by a factor $10^3$) as
functions of $a$ in $D=5$. Right:  $k_{BPS}$ and the ratios
$\frac{\Delta \, \mathcal{M}^2}{\mathcal{Q}_e^2}, \frac{a^2
\mathcal{M}^2}{\mathcal{D}^2}$.} \label{BPS5}
\end{figure}

\begin{figure}[ht]
\includegraphics[width=12cm]{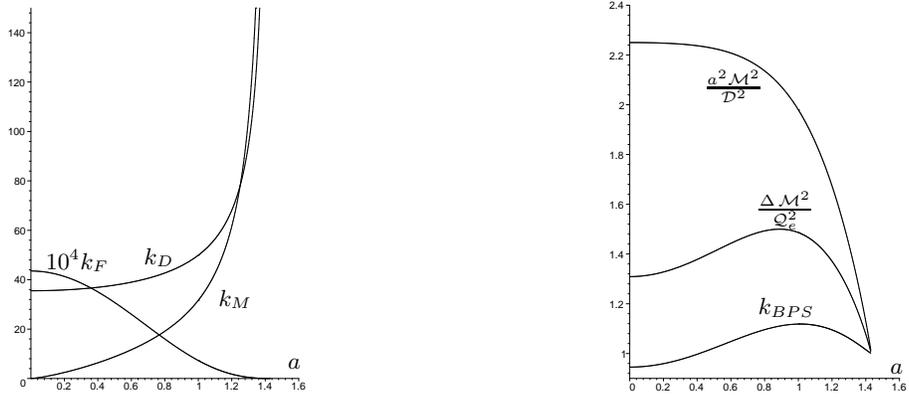}
\caption{Left: %ratios of mass, dilaton charge and the exponential
%of dilaton at spatial infinity  to the electrical charge $Q_e$, i.e.
$k_M(a), k_D(a)$ and $k_F(a)$ (multiplied by a factor $10^4$) in $D=6$.
Right: the ratios of $k_{BPS}$ and $\frac{\Delta \, \mathcal{M}^2}{\mathcal{Q}_e^2},
\frac{a^2 \mathcal{M}^2}{\mathcal{D}^2}$.}
\label{BPS6}
\end{figure}

%%%%%%%%%%%%%%%%%%%%%%%%%%%%%%%%%%%%%%%%%%%%%%%%%%%%%%%%%%%%%%%%%%%%%%
For $D = 5$ and $6$, the dynamics of the system is similar: the
asymptotically flat solutions exist up to the critical value of
dilaton coupling $a_{\rm cr}$, and in the limit $a \to a_{\rm cr}$
the BPS conditions are satisfied, namely $k_{BPS} \to 1$ as shown in
Figs.~\ref{BPS5} and \ref{BPS6}.

The critical value of dilaton coupling for $D=5$ is found to be
$a_{\rm cr}^{D=5} = 0.872422988$. The corresponding string value
(\ref{astring}) is smaller: $a^{D=5}_{\rm str}=0.816496580$, so
asymptotically flat stretched dilatonic black holes exist in five
dimensions.

For $D=6$ we obtain $a_{\rm cr}^{D=6} = 1.432881972$, the
corresponding string value (\ref{astring}) being  $a^{D=6}_{\rm
str}=0.707106781)$. We observe that while the critical dilaton
coupling is increasing with $D$, the string value (\ref{astring}) is
decreasing, so we expect that for $D>4$ we will have always
$a_{\rm str} < a_{\rm cr}$).

\begin{figure}[ht]
\includegraphics[width=12cm]{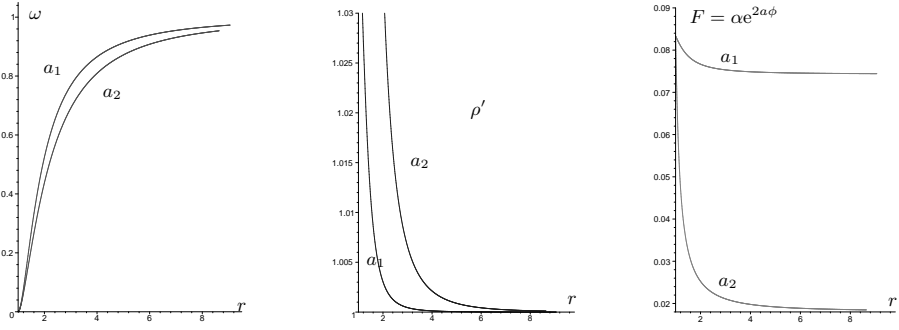}
\caption{Radial dependence of metric functions $\omega, \rho'$ and
the dilaton exponential $F$ for asymptotically flat black holes ($a <
a_{\rm cr}^{D=5}$) with dilaton couplings $a_1 = 0.3$ and $a_2 =
0.8$ in $D=5$.} \label{AF5}
\end{figure}

\begin{figure}[ht]
\includegraphics[width=12cm]{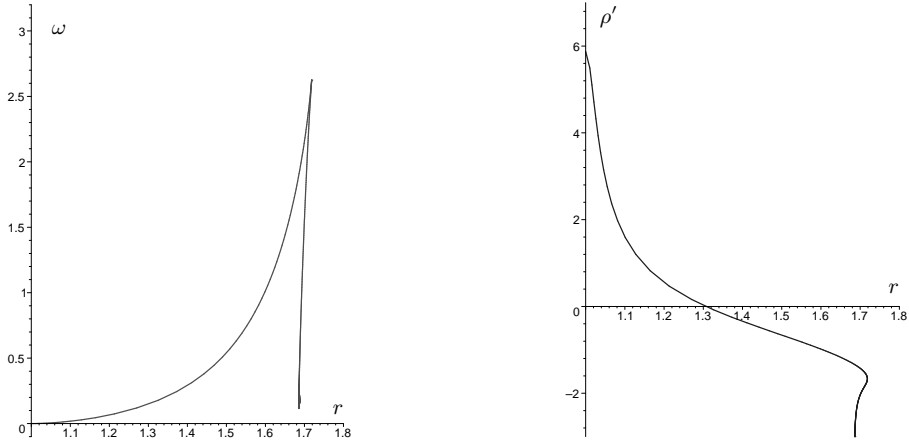}
\caption{Radial dependence of metric functions $\omega, \rho'$ for
singular solutions ($a > a_{\rm cr}^{D=5}$) for dilaton couplings $a
= 0.9$ in $D=5$.} \label{NAF5}
\end{figure}

\begin{figure}[ht]
\includegraphics[width=12cm]{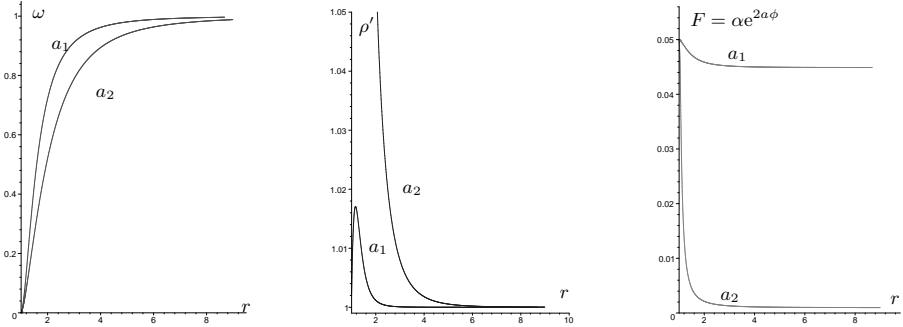}
\caption{Radial dependence of metric functions $\omega, \rho'$
and an exponential of dilaton field $F$ for asymptotical flat black holes
($a < a_{\rm cr}^{D=6}$) of dilaton couplings $a_1 = 0.4$, $a_2 = 1.4$
in $D=6$.}
\label{AF6}
\end{figure}

\begin{figure}[ht]
\includegraphics[width=12cm]{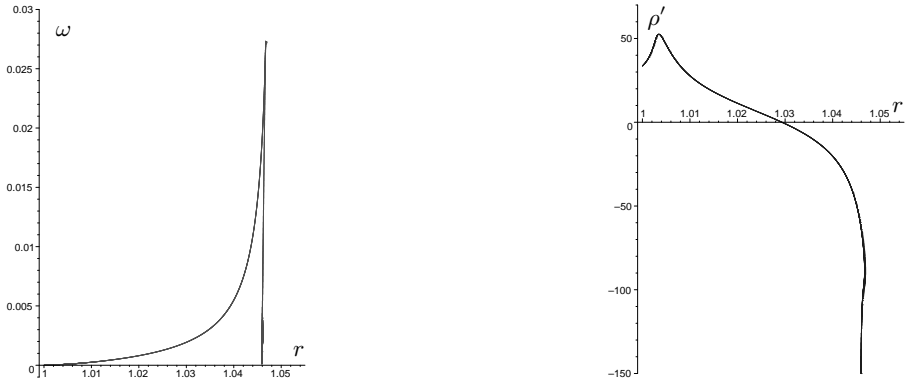}
\caption{Radial dependence of metric functions $\omega, \rho'$
for singular solutions ($a > a_{\rm cr}^{D=6}$) of dilaton coupling $a = 1.5$
in $D=6$.}
\label{NAF6}
\end{figure}

The metric functions and the dilaton exponential $F$ for
asymptotically flat black holes are given in Fig.~\ref{AF5} for
$a = 0.3$ and $0.8$ (which are smaller than the critical value in
$D = 5$), and in Fig.~\ref{AF6} for $a = 0.4$ and $1.4$ in $D = 6$. Those
for $a$ larger than the critical value are displayed in
Fig.~\ref{NAF5} in $D = 5$ and in Fig.~\ref{NAF6} in $D = 6$.

%%%%%%%%%%%%%%%%%%%%%%%%%%%%%%%%%%%%%%%%%%%%%%%%%%%%%%%%%%%%%%%%%%%%%%
\subsection{$D = 7$}
\label{d7}
\begin{figure}[ht]
\includegraphics[width=10cm]{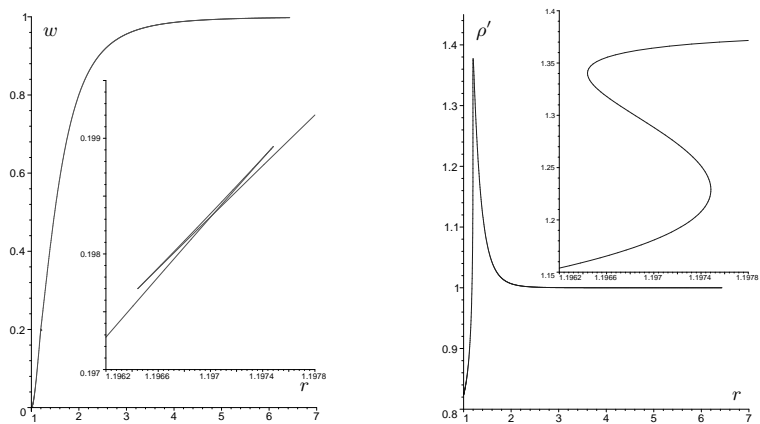}
\caption{Dependence of metric functions $w$ and $\rho'$ on radial coordinate
for $a=1.8$ in $D=7$ with two turning points.
(Insets describe properties of solution nearby turning points).}
\label{AF7a}
\end{figure}

\begin{figure}[ht]
\includegraphics[width=10cm]{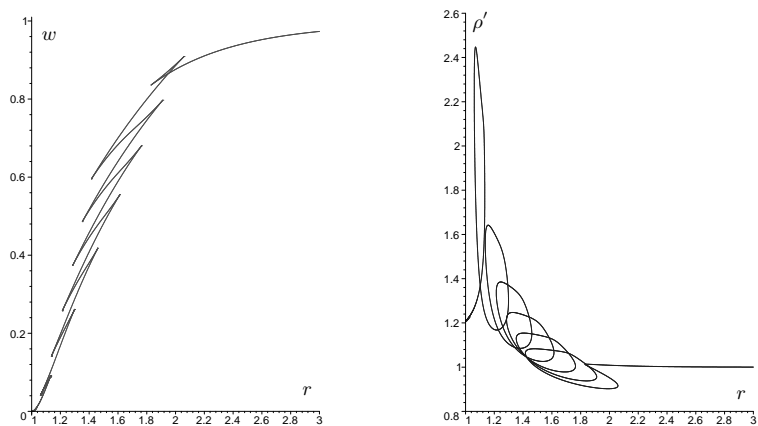}
\caption{Dependence metric functions $w$ and $\rho'$ of radial coordinate for
$a=1.81$ in $D=7$. There are fourteen turning points.}
\label{AF7b}
\end{figure}

\begin{figure}[ht]
\includegraphics[width=12cm]{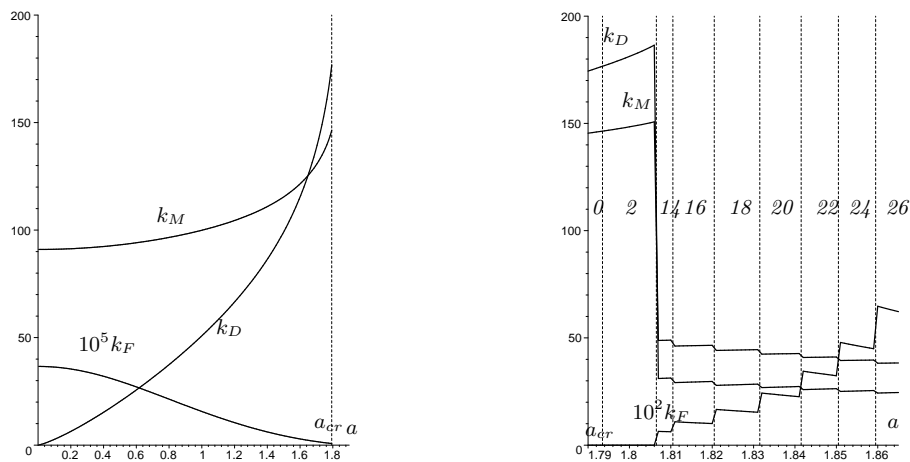}
\caption{Left: %Ratios of mass, dilaton charge and the exponential
%of dilaton at spatial infinity to the electrical charge $Q_e$, i.e.
$k_M(a), k_D(a)$ and $k_F(a)$ in the region before
formation of turning point in $D=7$.
Right: same quantities after formation of turning points
(the number of turning points is denoted by italic font).}
\label{k7}
\end{figure}

%%%%%%%%%%%%%%%%%%%%%%%%%%%%%%%%%%%%%%%%%%%%%%%%%%%%%%%%%%%%%%%%%%%%%%
The critical value for the dilaton coupling $a^{D=7}_\mathrm{cr} =
1.793909999$, the heterotic string value being $a^{D=7}_{\rm str} =
0.632455532$. As in lower dimensions, the critical value
corresponds to the appearance of the first cusp (turning point) in the
solution. The novel feature in $D = 7$ is that after the right turning
point the left one appears, and the solution can be extended
along the lines of~\cite{Alexeev:1997ua}. Using the same procedure
one can extend the solution to an asymptotically flat one, as shown
in Fig.~\ref{AF7a}. With further increasing dilaton coupling the
number of pairs of the turning points increases, so the
asymptotically flat extended solutions look as shown in
Fig.~\ref{AF7b}. The global parameters change in step-function-like manner
each time when one new turning point is created, see
Fig.~\ref{k7}. However, the extended solution cannot be considered
as true black hole solution, since geodesics, as we have shown in
Sec.~V~B, cannot be continued smoothly through the cusp
singularities. Therefore we have to consider the critical value of
the dilaton coupling in $D = 7$ as the true boundary of the range of
$a$.

\begin{figure}[ht]
\includegraphics[width=12cm]{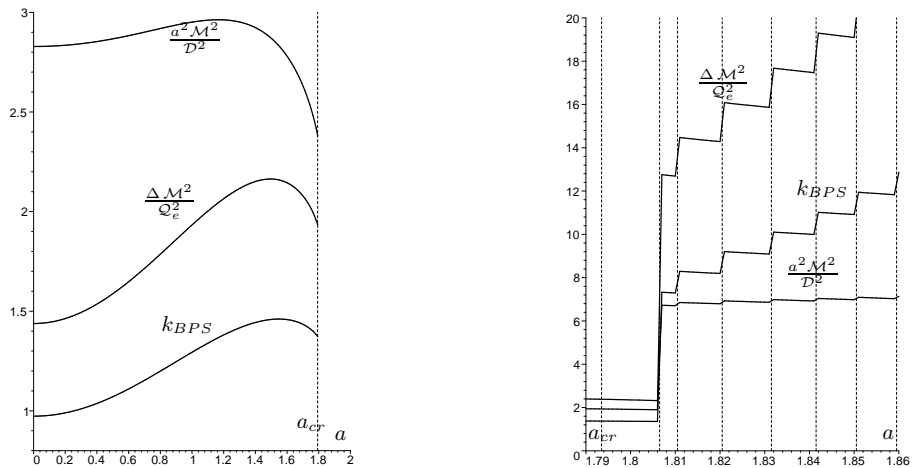}
\caption{Left: Ratios of $k_{BPS}$ and
$\frac{\Delta \, \mathcal{M}^2}{\mathcal{Q}_e^2},
\frac{a^2 \mathcal{M}^2}{\mathcal{D}^2}$ in the region before formation of
turning point in $D=7$.
Right: those after formation of turning points (the number of turning points
is denoted by italic font).}
\label{BPS7}
\end{figure}

In the case $D = 4$~\cite{Chen:2006ge} it was observed that on both
boundary of the dilaton coupling $a \to 0$ and $a \to
a_\mathrm{cr}$, the BPS condition of the EMD theory is saturated.
This can be understood as indication that the GB term is decoupled
in these two limits. Indeed, for $a=0$ it is obvious, while in the
limit $a \to a_\mathrm{cr}$ the mass tends to infinity, in which
case the Einstein term turns out to be dominant. In higher
dimensions situation is different. For $D \ge 5$ decoupling of the
dilaton $a=0$ does not switch off the GB term; instead we have to deal with
Wiltshire solutions of the EMGB theory. So the BPS saturation is not
expected for $a=0$, and this is confirmed by numerical calculations. For
$D=5, 6$ the BPS condition still holds on the right boundary of the
dilaton coupling $a \to a_\mathrm{cr}$ (see Figs.~\ref{BPS5},
\ref{BPS6}). However, in $D=7$, the BPS condition does not hold
anymore (see Fig.~\ref{BPS7}). This means that the GB term does not
decouple in the limit $a \to a_\mathrm{cr}$ as in lower dimensions.

\begin{figure}[ht]
\includegraphics[width=10cm]{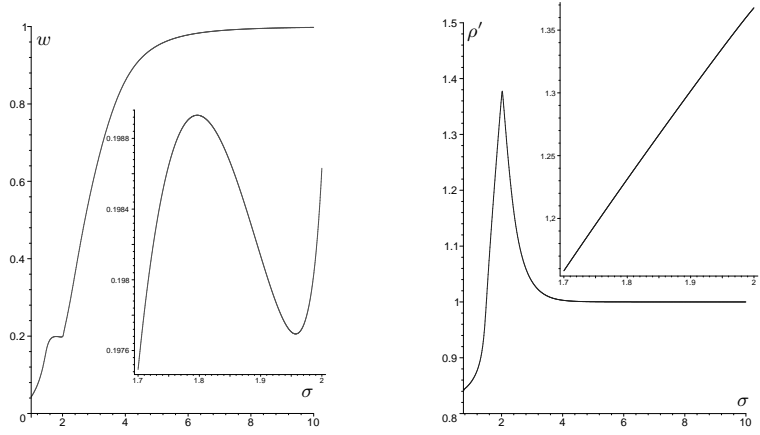}
\caption{Dependence metric functions
$w$ and $\rho'$ of parameter $\sigma$ for $a = 1.8$ in $D=7$ with two turning points.
(insets describe properties of solution nearby turning points).}
\label{B8}
\end{figure}

\begin{figure}[ht]
\includegraphics[width=10cm]{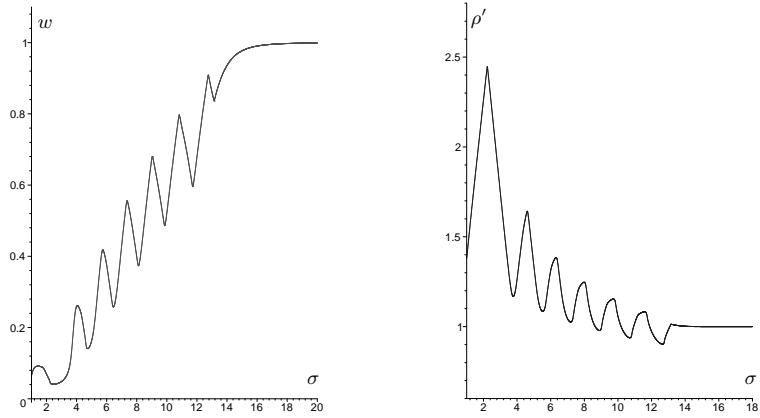}
\caption{
Dependence metric functions $w$ and $\rho'$ of parameter $\sigma$ for $a=1.81$
in $D=7$ with fourteen turning points.}
\label{B9}
\end{figure}

\begin{figure}[ht]
\includegraphics[width=4cm]{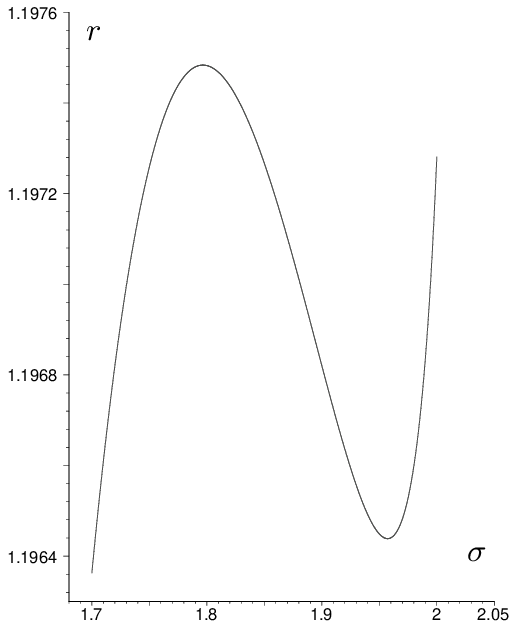}
\caption{Dependence original radial
coordinate $r$ of parameter $\sigma$ for $a=1.8$ in $D=7$ with two turning points.}
\label{B10}
\end{figure}

%%%%%%%%%%%%%%%%%%%%%%%%%%%%%%%%%%%%%%%%%%%%%%%%%%%%%%%%%%%%%%%%%%%%%%
\subsection{$D = 8, 9, 10$}
\label{d8}
\begin{figure}[ht]
\includegraphics[width=12cm]{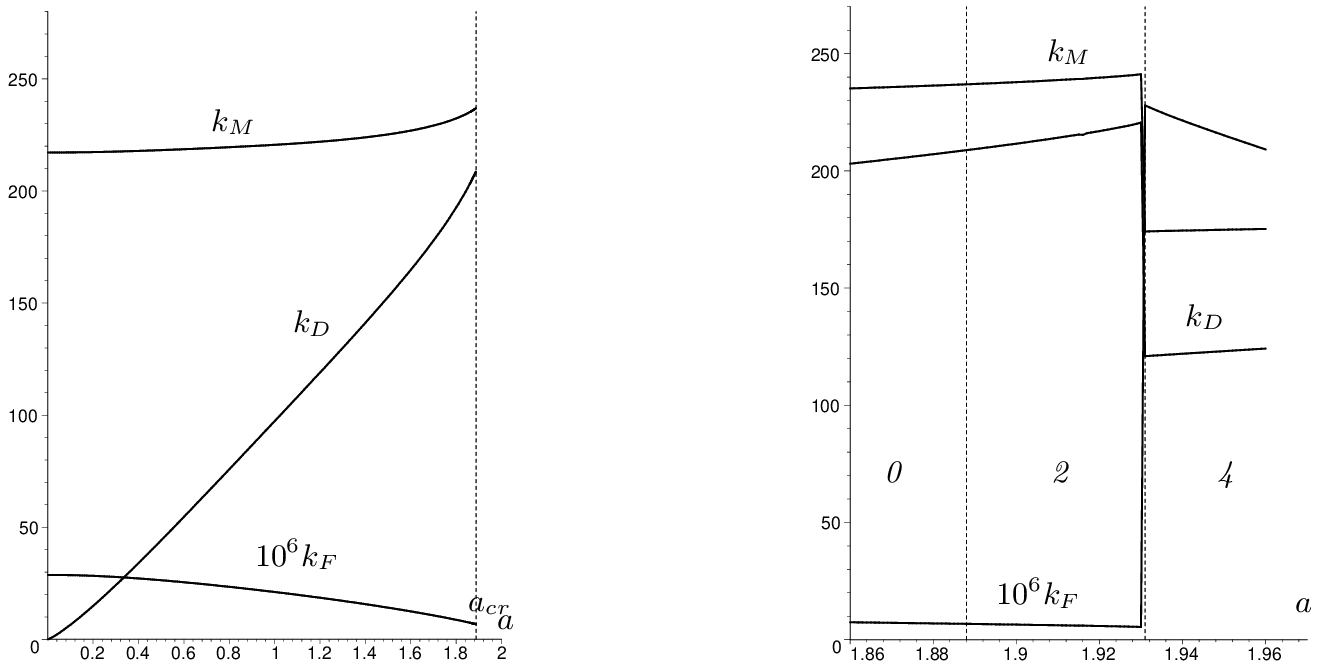}
\caption{%($D=8$) Ratios of mass, dilaton charge and the exponential
%of dilaton at spatial infinity to the electrical charge $Q_e$, i.e.
Left: $k_M(a), k_D(a)$ and $k_F(a)$ in $D=8$ in the region before
formation of turning point. Right: those after formation of turning points.}
\label{B11}
\end{figure}

\begin{figure}[ht]
\includegraphics[width=12cm]{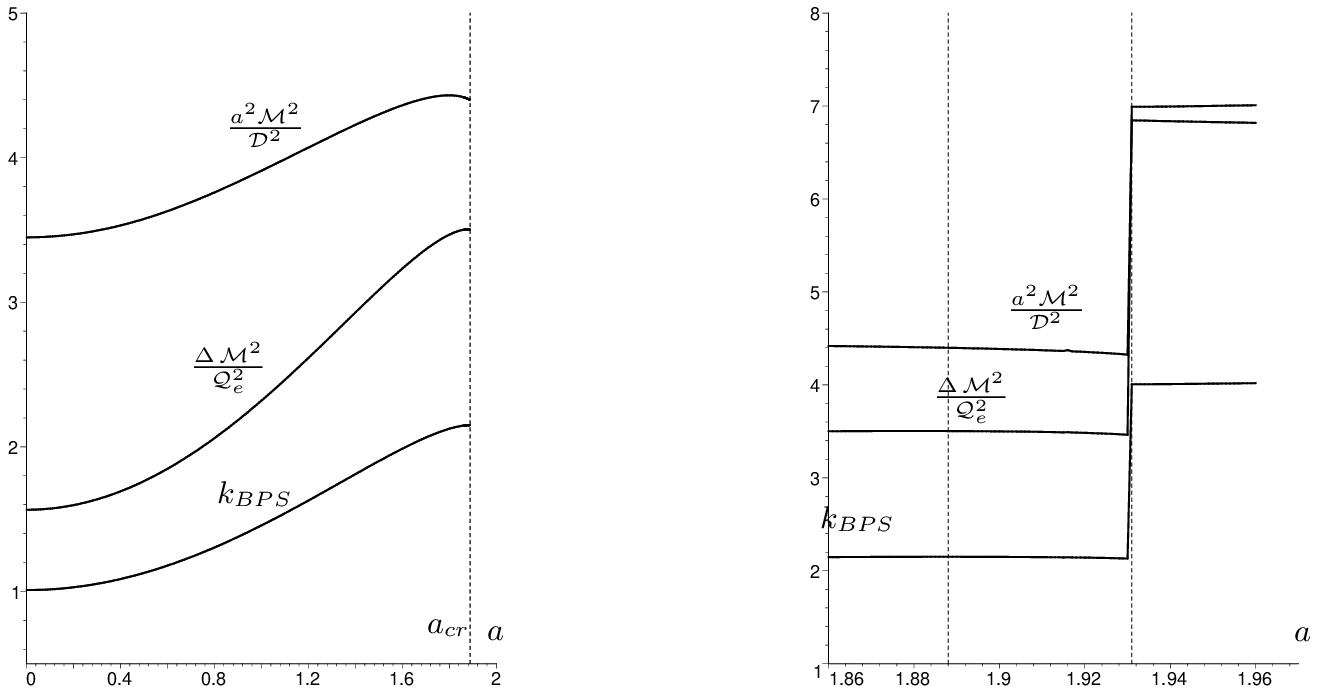}
\caption{%($D=8$)
Left: Ratios of $k_{BPS}$ and $\frac{\Delta \,
\mathcal{M}^2}{\mathcal{Q}_e^2}, \frac{a^2 \mathcal{M}^2}{\mathcal{D}^2}$
in$D=8$ in the region before formation of turning point.
Right: those after formation of turning points.}
\label{r8}
\end{figure}

\begin{figure}[ht]
\includegraphics[width=12cm]{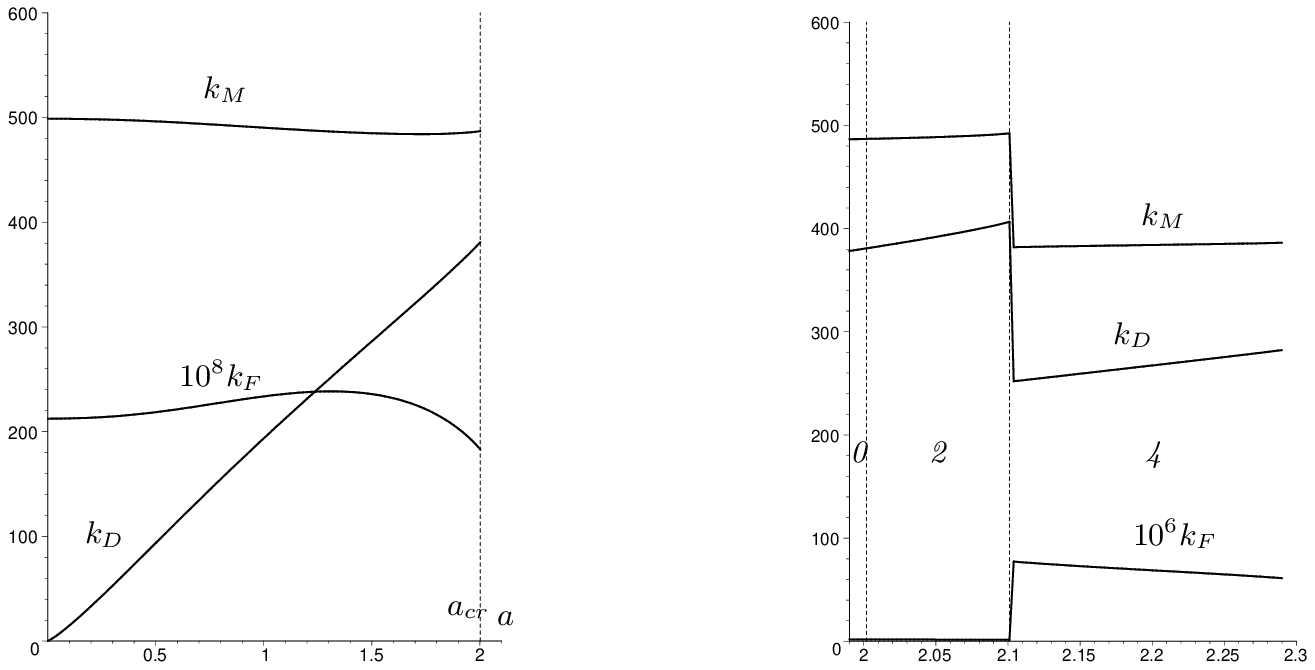}
\caption{%($D=9$) Ratios of mass, dilaton charge and the exponential of dilaton
%at spatial infinity to the electrical charge $Q_e$, i.e.
Left: $k_M(a), k_D(a)$ and $k_F(a)$ in $D=9$ in the region before formation of
turning point.
Right: those after formation of turning points.}
\label{B12}
\end{figure}

\begin{figure}[ht]
\includegraphics[width=12cm]{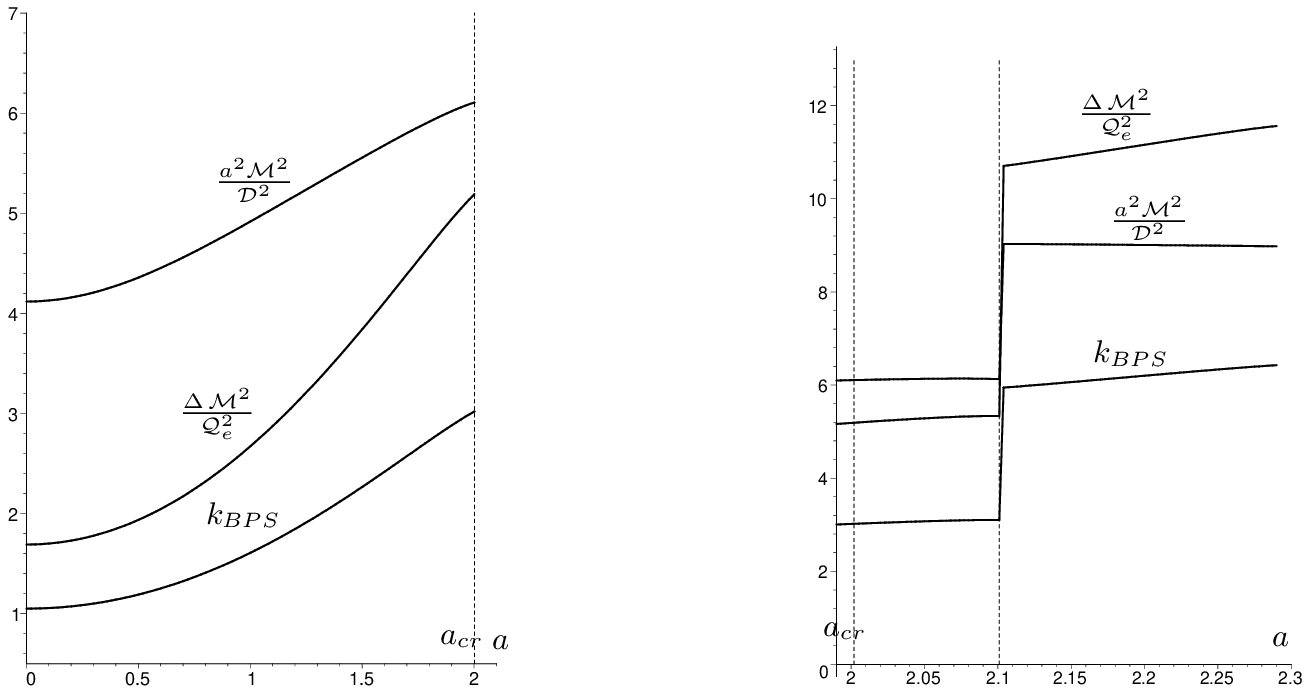}
\caption{%($D=9$)
Left: Ratios of $k_{BPS}$ and $\frac{\Delta \,
\mathcal{M}^2}{\mathcal{Q}_e^2}, \frac{a^2 \mathcal{M}^2}{\mathcal{D}^2}$ in $D=9$
in the region before formation of turning point.
Right: those after formation of turning points.}
\label{r9}
\end{figure}

\begin{figure}[ht]
\includegraphics[width=12cm]{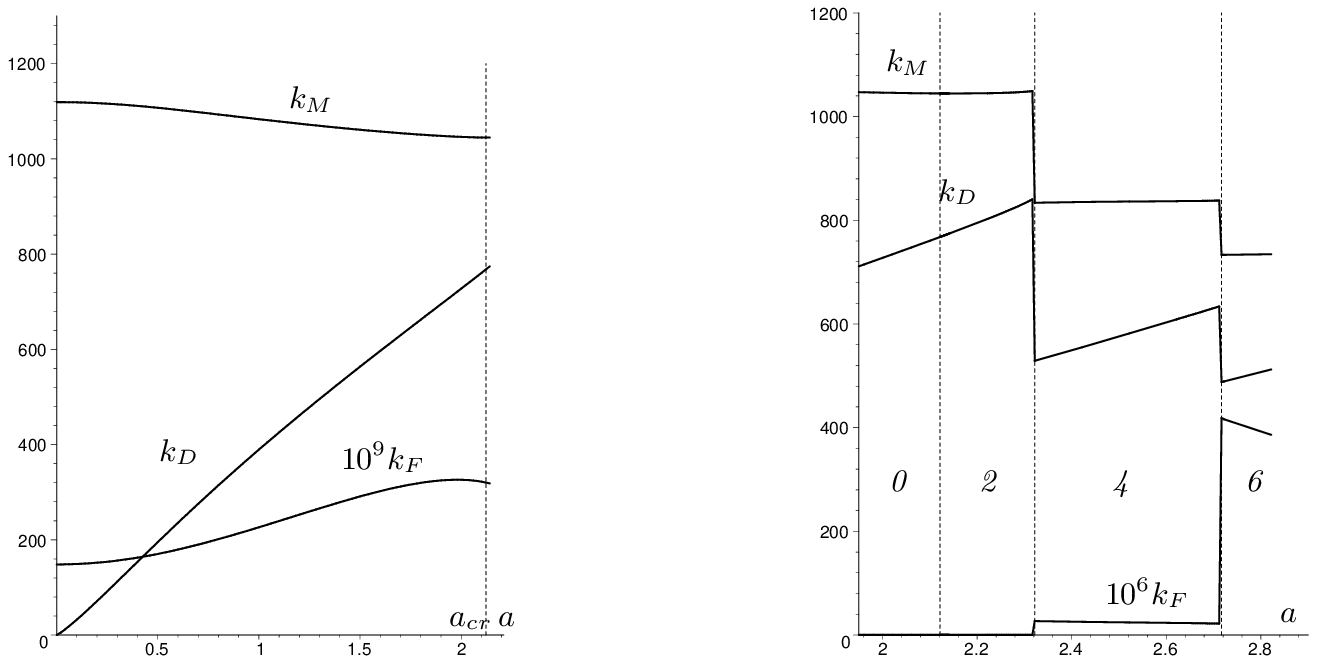}
\caption{%($D=10$) Ratios of mass, dilaton charge and the exponential
%of dilaton at spatial infinity to the electrical charge $Q_e$, i.e.
Left: $k_M(a), k_D(a)$ and $k_F(a)$ in $D=10$ in the region before
formation of turning point. Right: those after formation of turning
points.} \label{B13}
\end{figure}

\begin{figure}[ht]
\includegraphics[width=12cm]{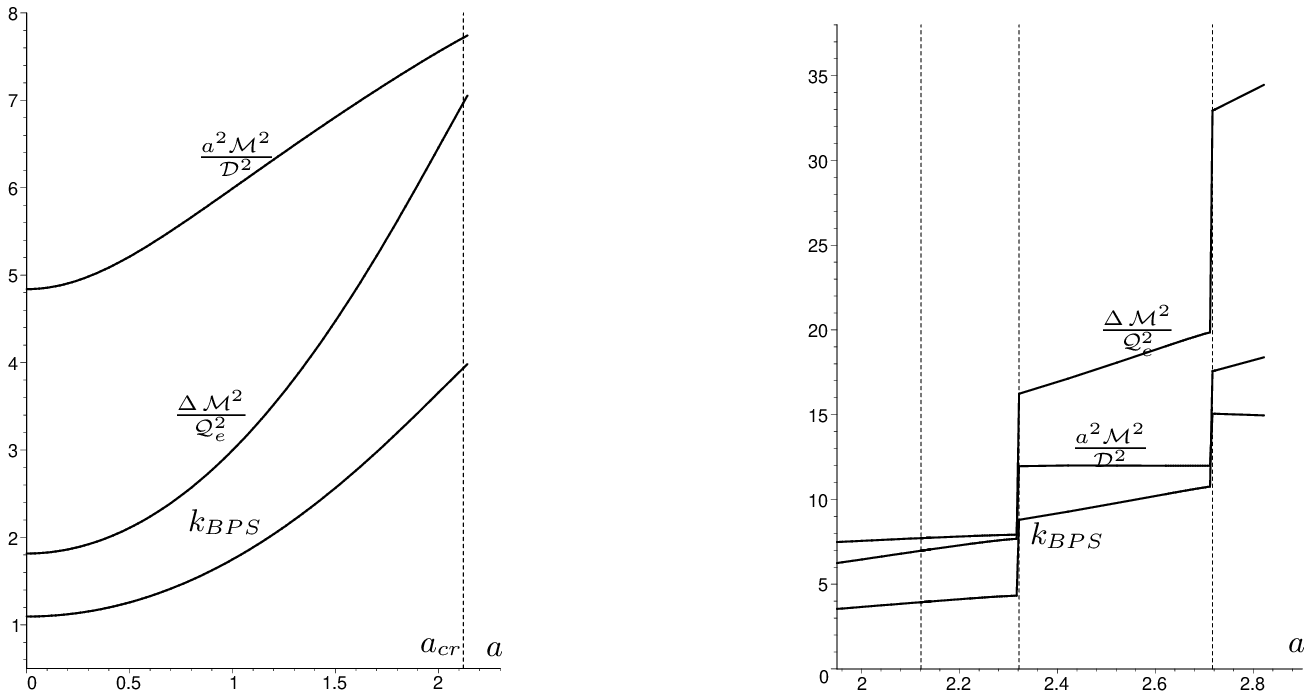}
\caption{%($D=10$)
Left: Ratios of $k_{BPS}$ and $\frac{\Delta \,
\mathcal{M}^2}{\mathcal{Q}_e^2}, \frac{a^2
\mathcal{M}^2}{\mathcal{D}^2}$ in $D=10$ in the region before
formation of turning point. Right: those after formation of turning
points.} \label{r10}
\end{figure}

%%%%%%%%%%%%%%%%%%%%%%%%%%%%%%%%%%%%%%%%%%%%%%%%%%%%%%%%%%%%%%%%%%%%%%
Properties similar to those in the case $D=7$ were observed for
higher dimensional solutions $D=8, 9, 10$. The critical values of
dilaton coupling are $a^{D=8}_\mathrm{cr} = 1.887653885$
($a^{D=8}_\mathrm{str} = 0.577350269$), $a^{D=9}_\mathrm{cr} =
2.002906751$ ($a^{D=9}_\mathrm{str} = 0.534522483$)  and
$a^{D=10}_\mathrm{cr} = 2.121748877$ ($a^{D=10}_\mathrm{str} = 0.5$). The numerical results
are presented in Figs.~\ref{B11}, \ref{r8}, \ref{B12}, \ref{r9}, \ref{B13}, \ref{r10}.
The BPS condition is not fulfilled on both boundaries of $a$.
Supercritical solutions can be formally
continued to infinity (as asymptotically flat) through the cusps which are
met in pairs like in the case of $D=7$. However we do not qualify them as
physical black holes for the reasons explained in Sec.~V.

%%%%%%%%%%%%%%%%%%%%%%%%%%%%%%%%%%%%%%%%%%%%%%%%%%%%%%%%%%%%%%%%%%%%%%
%%%%%%%%%%%%%%%%%%%%%%%%%%%%%%%%%%%%%%%%%%%%%%%%%%%%%%%%%%%%%%%%%%%%%%
\section {Conclusions}
\label{concl}
%%%%%%%%%%%%%%%%%%%%%%%%%%%%%%%%%%%%%%%%%%%%%%%%%%%%%%%%%%%%%%%%%%%%%%
%%%%%%%%%%%%%%%%%%%%%%%%%%%%%%%%%%%%%%%%%%%%%%%%%%%%%%%%%%%%%%%%%%%%%%
Here we summarize our findings. First, we have constructed explicit
local solution of the EMDGB static extremal black holes in the
vicinity of the horizon and calculated the corresponding entropies.
The ratios of entropy to the Hawking-Bekenstein one $A/4$ increases
from $1$ for $D = 4, 5$ to $41/13$ for $D = 10$. The entropy does not depend
on the dilaton coupling. Contrary to this, the asymptotic behavior
of the solutions crucially depend on dilaton coupling and
asymptotically flat black holes exist only for $a < a_{\rm cr}$. The
critical value of the dilaton coupling depend on $D$ and increases
with $D$. For $D = 4$, $a_{\rm cr}$ is smaller than the heterotic
string value, therefore no stretched black holes exist in the
effective heterotic theory. In contrast, for $D \ge 5$ the heterotic
values of $a$ lie inside the allowed region. Numerical solutions for
asymptotically flat black holes are constructed for $4 \le D \le 10$.
We investigated the ratios of the mass, dilaton charge and electric
charges which show the degree of deviation form the BPS bounds in
the absence of GB term as functions of the dilaton coupling. It is
observed that for $D < 5$ the BPS bound is saturated near the
threshold value $a_{\rm cr}$, thus demonstrating that the
contribution of the GB term is effectively small there. For larger
$D$ such a behavior was not observed, indicating that the GB term
remains important on the boundary.

The failure to reach the flat asymptotic in numerical integration
manifests itself as emergence of turning points of the radial
variable in which the scalar curvature has very mild
divergence. The solutions then exhibit typical cusp-shaped behavior.
It was suggested before that these turning points should be passed by
changing the integration variable in a suitable way so that the
solution can be continued through these singularities. We have found
that in dimensions $D \ge 7$ the turning points comes in pairs, and
the solution can be formally extended to the flat asymptotic.
However an inspection of radial geodesics reveals that
they cannot be analytically continued through cusp singularities,
so we do not believe that continuation of numerical solutions
through the cusps is physically meaningful.

%%%%%%%%%%%%%%%%%%%%%%%%%%%%%%%%%%%%%%%%%%%%%%%%%%%%%%%%%%%%%%%%%%%%%%
\section*{Acknowledgments}
%%%%%%%%%%%%%%%%%%%%%%%%%%%%%%%%%%%%%%%%%%%%%%%%%%%%%%%%%%%%%%%%%%%%%%
CMC is grateful to the AEI, Postdam for its hospitality in the early stages of this work.
The work of CMC and DGO was supported by the National Science Council of the
R.O.C. under the grant NSC 96-2112-M-008-006-MY3 and in part by the
National Center of Theoretical Sciences (NCTS).
The work of DG was supported by the RFBR under the project 08-02-01398-a.
The work of NO was supported in part by the Grant-in-Aid for
Scientific Research Fund of the JSPS Nos. 20540283, and also by the
Japan-U.K. Research Cooperative Program.

\begin{appendix}
%%%%%%%%%%%%%%%%%%%%%%%%%%%%%%%%%%%%%%%%%%%%%%%%%%%%%%%%%%%%%%%%%%%%%%
\section{Geometric Quantities for Spherical Symmetric Metric}
%%%%%%%%%%%%%%%%%%%%%%%%%%%%%%%%%%%%%%%%%%%%%%%%%%%%%%%%%%%%%%%%%%%%%%
This appendix gives detail geometric quantities associated with the following
spherical symmetric metric in and dimensions $D$
\begin{equation}
ds^2 = - \mathrm{e}^{2 u(r)} dt^2 + \mathrm{e}^{2 v(r)} dr^2
+ \mathrm{e}^{2 w(r)} d\Omega_{D-2, k}^2,
\end{equation}
where $k$ denotes the spatial curvature. The Riemann and Ricci tensors have
the following components
\begin{eqnarray}
R_{trtr} &=& \mathrm{e}^{2u} (u'' + u'^2 - u' v'),
\nonumber\\
R_{tatb} &=& \mathrm{e}^{2u - 2v + 2w} u' w' \; g_{ab},
\nonumber\\
R_{rarb} &=& - \mathrm{e}^{2w} (w'' + w'^2 - v' w') \; g_{ab},
\nonumber\\
R_{acbd} &=& \left( - \mathrm{e}^{4w - 2v} w'^2 + k \, \mathrm{e}^{2w} \right)
(1 - \delta_{ac} \delta_{bd}) \; g_{ab} g_{cd},
\nonumber\\
R_{tt} &=& \mathrm{e}^{2u - 2v} (u'' + u' H'),
\nonumber\\
R_{rr} &=& - (u'' + u'^2 - u' v') - (D - 2) (w'' + w'^2 - v' w'),
\nonumber\\
R_{ab} &=& \left[ - \mathrm{e}^{2w - 2v} (w'' + w' H') + k (D - 3) \right] g_{ab},
\end{eqnarray}
and then the scalar curvature, Ricci square ($R_{\mu\nu}^2 = R_{\mu\nu} R^{\mu\nu}$)
and Riemann square ($R_{\alpha\beta\mu\nu}^2 =
R_{\alpha\beta\mu\nu} R^{\alpha\beta\mu\nu}$) are
\begin{eqnarray}
R &=& - \mathrm{e}^{-2 v} \left[ 2 u'' + u' H' + u'^2 - u' v'
+ D^2_2 ( 2 w'' + w' H' + w'^2 - v' w' ) \right] + k D^2_3 \, \mathrm{e}^{-2 w},
\nonumber\\
&=& - \mathrm{e}^{-2 v} \left[ 2 (u'' + u'^2 - u' v') + 2 D^2_2 (w''
+ u' w' - v' w' + w'^2) + D^2_3 \left( w'^2 - k \mathrm{e}^{2v - 2w} \right) \right],
\nonumber\\
R_{\mu\nu}^2 &=& \mathrm{e}^{-4 v} (u'' + u' H')^2
+ \mathrm{e}^{-4 v} [ u'' + u'^2 - u' v' + D^2_2 ( w'' + w'^2 - v' w') ]^2
\nonumber\\
&+& D^2_2 \left[ \mathrm{e}^{-2 v} (w'' + w' H') - k D^3_3 \,
\mathrm{e}^{-2 w} \right]^2,
\\
R_{\alpha\beta\mu\nu}^2 &=& 4 \, \mathrm{e}^{-4 v} (u'' + u'^2 - u' v')^2
+ 4 D^2_2 \, \mathrm{e}^{-4 v} u'^2 w'^2 + 4 D^2_2 \, \mathrm{e}^{-4 v}
( w'' + w'^2 - v' w')^2
\nonumber\\
&+& 2 D^2_3 \left( \mathrm{e}^{-2 v} w'^2 - k \, \mathrm{e}^{-2 w} \right)^2, \nonumber
\end{eqnarray}
where $H$ is defined
\begin{equation}
H = u - v + (D - 2) w,
\end{equation}
and the following notation is used
\begin{equation}
D^m_n = (D - m)_n = (D - m) (D - m - 1) \cdots (D - n), \qquad n \ge m.
\end{equation}
The GB combination is
\begin{eqnarray}
{\cal L}_{\rm GB}
%&=& R^2 - 4 R_{\mu\nu} R^{\mu\nu} + R_{\alpha\beta\mu\nu} R^{\alpha\beta\mu\nu}
%\nonumber\\
%&=& 4 (D - 2)(D - 3) \left\{ (D-4) w'^2 (w'' + u' w' - v' w' + w'^2) + w'^2 (u'' + u'^2 - u' v') + 2 u' w' (w'' + w'^2 - w' v') \right\} \mathrm{e}^{- 4v}
%\nonumber\\
%&& - 4 k (D - 2)(D - 3) \left\{ u'' + u'^2 - u' v' + (D - 4) (w'' + w' u' - w' v' + w'^2) \right\} \mathrm{e}^{- 2v - 2w}
%\nonumber\\
%&& + (D-2) (D-3) (D-4) (D-5)  \left( w'^2 \, \mathrm{e}^{- 2v} - k \, \mathrm{e}^{- 2w} \right)^2.
%\\
&=& D^2_3 \, \mathrm{e}^{- 4 v} \biggl\{ D^4_5 \left( w'^2 - k \,
\mathrm{e}^{2v - 2w} \right)^2 + 8 u' w' (w'' + w'^2 - w' v')
\nonumber\\
&+& 4 \left[ u'' + u'^2 - u' v' + D^4_4 (w'' + u' w' - v' w' + w'^2) \right]
\left( w'^2 - k \, \mathrm{e}^{2v - 2w} \right) \biggr\}
\nonumber\\
&=& 4 D^2_3 \, \mathrm{e}^{- u - v - 2w} \left[ \mathrm{e}^{u - 3v + 2w} u'
\left( w'^2 - k \, \mathrm{e}^{2v - 2w} \right) \right]'
\nonumber\\
&+& 4 D^2_4 \, \mathrm{e}^{- 4 v} (w'' + u' w' - v' w' + w'^2)
\left( w'^2 - k \, \mathrm{e}^{2v - 2w} \right)
\nonumber\\
&+& D^2_5 \, \mathrm{e}^{- 4 v} \left( w'^2 - k \, \mathrm{e}^{2v - 2w} \right)^2.
\end{eqnarray}
One can easy check that, in four dimension, the GB term of $\sqrt{-g} {\cal L}_{\rm GB}$
is a total derivative.
%\begin{equation}
%{\cal L}_{\rm GB} = 8 \, \mathrm{e}^{-u - v - 2w} \left[ \mathrm{e}^{u - 3v + 2w}u' \left( w'^2 - k \, \mathrm{e}^{2v - 2w} \right) \right]'.
%\end{equation}
For the gauge choice of coordinates
\begin{equation}
u = - v = \frac12 \ln\omega, \qquad w = \ln\rho,
\end{equation}
the relevant quantities become
\begin{eqnarray}
R &=& - \rho^{-2} \left[ \omega'' \rho^2 + 2 D^2_2 \rho (\omega \rho''
+ \omega' \rho') + D^2_3 (\omega \rho'^2 - k) \right],
\nonumber\\
&=& \rho^{2-D} \left[ - \left( \omega' \rho^{D-2} + 2 D^2_2 \omega \rho'
\rho^{D-3} \right)' + D^2_2 \rho' (\omega \rho^{D-3})' + D^2_3 \rho^{D-4} k \right],
\\
{\cal L}_{\rm GB} &=& D^2_3 \rho^{-4} \left\{ 2 \omega' \rho' \rho^2 (2 \omega \rho''
+ \omega' \rho') + 2 \rho [\omega'' \rho + 2 D^4_4 (\omega \rho')']
(\omega \rho'^2 - k) + D^4_5 (\omega \rho'^2 - k)^2 \right\},
\nonumber\\
&=& 2 D^2_3 \rho^{-2} [\omega' (\omega \rho'^2 - k)]' + 4 D^2_4 \rho^{-3}
(\omega \rho'' + \omega' \rho') (\omega \rho'^2 - k) + D^2_5 \rho^{-4}
(\omega \rho'^2 - k)^2.
\end{eqnarray}

\end{appendix}

%%%%%%%%%%%%%%%%%%%%%%%%%%%%%%%%%%%%%%%%%%%%%%%%%%%%%%%%%%%%%%%%%%%%%%

\end{document}